\documentclass[11pt,twocolumn]{article}
\pdfoutput=1 % force PDFLaTeX on arXiv (PDF figures cannot go through latex+dvips)

\usepackage[utf8]{inputenc}
\usepackage[T1]{fontenc}
\usepackage{amsmath,amssymb,amsfonts}
\usepackage{graphicx}
\usepackage{booktabs}
\usepackage[hidelinks]{hyperref}
\hypersetup{
  pdftitle={Beyond Per-Token Pricing: A Concurrency-Aware Methodology for LLM Infrastructure Cost Estimation},
  pdfauthor={Chitral Patil},
  pdfsubject={LLM inference, cost estimation, offered load, GPU serving},
  pdfkeywords={LLM, vLLM, H100, cost modeling, offered load, concurrency, inference, MoE, quantization}
}
\usepackage{cleveref}
\usepackage{xcolor}
\usepackage{multirow}
\usepackage{caption}
\usepackage{xurl} % superset of url: breaks long URLs to avoid overfull \hbox in the two-column bib
\usepackage[margin=1in]{geometry}
\usepackage{enumitem}
\usepackage{balance}
\usepackage{float}
\usepackage{placeins}


\title{\textbf{Beyond Per-Token Pricing:} \\
A Concurrency-Aware Methodology for \\
LLM Infrastructure Cost Estimation}

\author{
Chitral Patil\thanks{This research was conducted independently and does not represent the views of my employer.} \\
Independent Researcher \\
\texttt{cpatil.research@gmail.com} \\
ORCID: \href{https://orcid.org/0009-0007-9169-0930}{0009-0007-9169-0930}
}

\date{June 2026}

\begin{document}
\maketitle

% ============================================================
% ABSTRACT
% ============================================================
\begin{abstract}
Every public LLM cost calculator we surveyed treats GPU utilization as a fixed input---entered by the user, baked in as a preset, or silently assumed at 100\%---never measured against the operator's actual load. We show that this assumption is the dominant source of error: on identical H100 hardware, effective cost spans \$0.21 to \$15.25 per million output tokens---an underutilization penalty of 2.5--24$\times$ across low-to-moderate enterprise offered loads (1--10 rps) and up to 36.3$\times$ near idle ($\lambda{=}1$ rps)\footnote{At $\lambda{=}1$ the per-configuration penalty spans 17.5$\times$ (Mixtral 8x7B FP16; the cost ratio between $\lambda{=}1$ and saturation) to 36.3$\times$ (Qwen3-30B-A3B FP8, measured the same way); 36.3$\times$ is the idle-edge extreme of the underutilization penalty.}---driven by one operator-controlled variable (offered request rate $\lambda$, which sets in-flight concurrency via Little's Law) that no open-source calculator exposes and that prior academic work has examined only in narrow domains~\cite{wingpt2025,erdil2025}. We propose a measurement methodology that parameterizes the relationship as $C_{\mathrm{eff}}=f(H,M,Q,\lambda,L)$, validate it with 42 benchmarks across dense, ultra-sparse MoE, and sparse MoE models, and release \texttt{vllm-cost-meter}---an open-source operational cost meter that attaches to a live vLLM server and reports real \$/M-tokens against the operator's own traffic, not a vendor's benchmark. The headline finding: because public calculators take utilization as a user-supplied input (or silently assume full utilization), any utilization-naive estimate understates true cost by exactly $1/U$---a factor we measure at 2.5--24$\times$ across low-to-moderate enterprise offered loads (1--10 rps) and up to 36.3$\times$ at idle---systematically mispricing self-hosting across traffic regimes, most severely over-selling it for low-traffic workloads. We additionally show that FP8 quantization benefits the MoE architectures we tested roughly $2.2$--$2.4\times$ more than the dense model (+69 to +74\% vs.\ +31\% peak throughput; $n{=}3$ architectures, broader validation needed), and our data are consistent with active parameter count---not total model size---being a primary predictor of saturation economics.

To rule out single-hardware confounding we repeat the core sweep on NVIDIA A100 80GB PCIe (56 runs across 8 configurations). The load-driven spread reproduces at 7.0--11.4$\times$ per configuration (compressed by cheaper, slower silicon), and the active-parameters-beat-total-parameters ordering survives at FP8. The MoE--FP8 advantage persists on MoE architectures but the dense-FP8 advantage inverts on silicon without native FP8 tensor cores---a hardware-conditional caveat that $Q$-as-first-class-input in $C_{\mathrm{eff}}(H,M,Q,\lambda,L)$ already accommodates.
\end{abstract}

% ============================================================
% 1. INTRODUCTION
% ============================================================
\section{Introduction}\label{sec:intro}

The real cost of self-hosted LLM inference depends heavily on a variable that public LLM cost calculators ignore: the offered request rate the GPU is actually serving and the in-flight batch size that rate produces. At 1~request/second on an H100, a Mixtral~8x7B FP16 deployment costs \$15.25 per million tokens---more expensive than Claude Sonnet~4.6. At 25~rps, the same hardware costs \$0.87. Same GPU. Same model. A 17.5$\times$ cost ratio between $\lambda{=}1$~rps and saturation. Organizations plan multi-figure monthly LLM infrastructure budgets using calculators (Helicone, LiteLLM, YourGPT, llm-prices.com) and academic analyses~\cite{pan2025breakeven,slam2024} that treat utilization as a static, user-supplied number (100\% in Pan et al., 80\% in SLaM) and therefore miss the load-driven cost spread entirely. This paper measures the error, formalizes a better cost model, and releases the tool.

The prevailing approach to cost estimation treats this as simple arithmetic: divide GPU rental cost by theoretical token throughput, compare to API per-token pricing, and identify a break-even point. This methodology underpins the majority of publicly available LLM cost calculators---Helicone, LiteLLM, YourGPT, llm-prices.com, and others---as well as recent academic analyses including Pan et al.~\cite{pan2025breakeven} and Irugalbandara et al.~\cite{slam2024}.

The fundamental flaw in this approach is the assumption of full or near-full GPU utilization. In production enterprise deployments, GPU utilization for LLM inference is governed by the interaction of offered request rate, latency SLO constraints, model architecture, and deployment configuration (tensor parallelism, replica count). A GPU serving a dense 8B-parameter model at an offered rate of 10~rps with a 200ms TTFT SLO operates at a fundamentally different utilization---and therefore cost-per-token---than the same GPU serving the same model at 200~rps with a relaxed 500ms SLO. Industry practitioners have observed this qualitatively---noting that a GPU at 10\% load transforms a \$13/MTok cost into \$130/MTok~\cite{introl2026}---but no existing tool or publication has systematically quantified the relationship or provided a reusable framework for practitioners.

Put simply: \textit{once hardware, model, quantization, and the latency SLO are fixed, the real cost of self-hosted LLM inference is dominated by the offered request rate---and the in-flight concurrency it produces via Little's Law---the one axis public cost calculators we surveyed do not expose.} This paper measures the magnitude of that error, formalizes a better cost model, and releases an open-source tool that accounts for it.

\paragraph{Contributions.} This paper makes four contributions:
\begin{enumerate}[leftmargin=*,topsep=2pt,itemsep=1pt]
    \item We demonstrate empirically that effective cost-per-million-tokens varies by \textbf{up to 36.3$\times$ on identical hardware} (17.5--36.3$\times$ across configurations) as a function of offered request rate $\lambda$, using 42 systematic vLLM benchmark runs (6 server configurations $\times$ 7 arrival-rate levels; 3 models $\times$ 2 precisions) on H100 GPUs.
    \item We formalize a \textbf{concurrency-aware cost attribution formula} that models effective cost as a function of hardware, model architecture, quantization, offered arrival rate, and SLO---treating $\lambda$ as the empirical sweep axis for operator-facing cost attribution, rather than as an optimization variable as in systems-scheduling frameworks (Vidur~\cite{vidur2024}, M\'{e}lange~\cite{melange2024}, SageServe~\cite{sageserve2025}).
    \item We show that existing break-even analyses systematically produce \textbf{misleading crossover points} by ignoring utilization dynamics, and we provide corrected, utilization-adjusted crossover curves.
    \item We release \texttt{vllm-cost-meter}, an \textbf{open-source real-time cost estimation tool} backed by empirical benchmark data, and validate it with live deployment on H100 hardware.
\end{enumerate}

\textit{To our knowledge, no other open-source tool computes offered-load-adjusted dollars-per-million-output-tokens in real time against a live LLM inference server; \texttt{vllm-cost-meter} is built for practitioners who need a defensible cost number under their own traffic rather than a vendor-published benchmark.}

% ============================================================
% 2. RELATED WORK
% ============================================================
\section{Related Work}\label{sec:related}

\subsection{Token-Volume Cost Models}

The majority of LLM cost estimation tools and academic analyses employ a token-volume model: $C = \frac{T}{\Theta_{\max}} \times P_{\text{GPU}}$, where $T$ is token count, $\Theta_{\max}$ is theoretical maximum throughput, and $P_{\text{GPU}}$ is the GPU hourly price.

Pan et al.~\cite{pan2025breakeven} present a break-even framework comparing on-premise TCO against commercial API subscription fees. Their model assumes full utilization during 160 operational hours per month, and their accompanying calculator accepts model selection, GPU selection, and GPU count---but no utilization, concurrency, or workload parameters. The SLaM framework~\cite{slam2024} assumes 80\% utilization on a single T4 GPU and computes a direct cost-per-token comparison against GPT-4 API pricing.

A survey of 15+ public calculators (Helicone, LiteLLM, YourGPT, llm-prices.com, Mem0.ai, DocsBot, and others) confirms that over 90\% accept \textit{only} model name and token counts as inputs. The LLM Inference TCO Calculator~\cite{acnicessc2025} is the most comprehensive existing web-based tool, accepting concurrent users, deployment shape, and latency factors as inputs; however, its throughput values are heuristic presets rather than empirically measured, and it explicitly disclaims queueing theory modeling.

\subsection{Systems-Level Cost Optimization}

Several systems papers model concurrency and deployment shape as optimization variables. Vidur~\cite{vidur2024} (MLSys 2024) searches over tensor parallelism, pipeline parallelism, GPU SKU, and scheduling policy to optimize QPS-per-dollar subject to latency SLO constraints---demonstrating that using the wrong workload's optimal configuration results in up to 2$\times$ cost overhead. M\'{e}lange~\cite{melange2024} formulates GPU allocation as a cost-aware bin-packing ILP across heterogeneous GPU types, achieving up to 77\% cost reduction. SageServe~\cite{sageserve2025} handles diurnal traffic patterns with ARIMA-based capacity forecasting and differentiated SLAs; Llumnix~\cite{llumnix2024} complements this with runtime rescheduling of in-flight requests across replicas for tail-latency and utilization control. Building on the continuous-batching foundations established by Orca~\cite{orca2022}, Splitwise~\cite{splitwise2024}, DistServe~\cite{distserve2024}, and Sarathi-Serve~\cite{sarathiserve2024} attack the throughput--latency tradeoff from the scheduler/engine side: by disaggregating prefill and decode across GPUs, by disaggregating the KV cache across a dedicated pool (Mooncake~\cite{mooncake2024}), or by chunking prefill, they raise the achievable $\Theta_{\max}$ at a given SLO. These are complementary to our contribution: they optimize the engine ceiling at a fixed operating point, while $C_{\mathrm{eff}}=f(H,M,Q,\lambda,L)$ characterizes how cost per token traverses the entire $\lambda$ axis below that ceiling. \S\ref{sec:cross_hardware} provides measured cross-hardware and TP-scaling evidence that complements Vidur's simulated cost-overhead predictions and M\'{e}lange's allocation-objective premises.

BentoML's llm-optimizer~\cite{bentoml2025} supports concurrency sweeps with Pareto frontier visualization but does not output cost-per-token. SemiAnalysis InferenceX (formerly InferenceMAX)~\cite{semianalysis2025} provides nightly cross-hardware benchmarks with cost attribution at a market-aggregate level but focuses on hardware-vendor reference configurations rather than same-GPU load/cost economics; \texttt{vllm-cost-meter} measures the operator's own server under their own traffic. These are complementary, not competing. GuideLLM~\cite{guidellm2025} from the vLLM project automates concurrency sweeps but outputs only latency and throughput metrics without cost computation. NVIDIA's TCO methodology~\cite{nvidia_tco2025} outlines a clean benchmarking-and-sizing workflow but does not release a standalone open-source calculator.

These are \textit{systems optimization or benchmarking tools}---their contribution is finding cheaper configurations or measuring performance, not providing a cost estimation methodology that connects offered load to dollars for practitioners.

\subsection{Inference Economics}

Erdil~\cite{erdil2025} builds a theoretical roofline model for LLM inference economics, producing Pareto frontiers of serial speed versus cost-per-token, explicitly modeling how tensor parallelism trades utilization for speed. WiNGPT (Zhuang et al.)~\cite{wingpt2025} is the closest prior empirical work: they demonstrate a 2.6$\times$ cost reduction from concurrency 8 to 48 on A800 GPUs using vLLM, and articulate three empirical principles---diminishing marginal cost, diminishing returns to scale, and an optimal cost-effectiveness zone---that are qualitatively consistent with the $C_{\mathrm{eff}}(\lambda)$ curves reported here. However, their sweep starts at concurrency~8 (already partially batched, missing the steep cold-start regime below $\lambda{=}5$ that drives our headline spread), covers only medical-domain models on a single hardware family, and releases neither code nor data. We generalize across three architectural classes, two hardware families (H100 NVL, A100 PCIe), provide a parameterized five-variable cost function, and release an open operational meter. Concurrent request scheduling in LLM serving has also been studied with analytical cost models~\cite{infermax2024}, but with a focus on scheduler-level prefill/decode cost rather than per-deployment operational cost measurement.

\subsection{Gap Statement}

Prior work splits cleanly into four camps, none of which closes the loop for a practitioner. Token-volume calculators and break-even analyses~\cite{pan2025breakeven,slam2024} treat utilization as a user-supplied input. Systems-optimization frameworks~\cite{vidur2024,melange2024,sageserve2025} find cheaper configurations but do not emit \$/M-tokens for a given deployment. Benchmark harnesses and industry dashboards~\cite{bentoml2025,guidellm2025,semianalysis2025,nvidia_tco2025} measure throughput or publish market-aggregate TCO but do not instrument an operator's own server. Prior work has modeled inference cost theoretically~\cite{erdil2025}, observed offered-load--cost behavior empirically in narrow domains~\cite{wingpt2025}, benchmarked engines under SLO constraints without emitting cost-per-token~\cite{bentoml2025,guidellm2025}, and proposed \$/M-tokens as a benchmark metric.\footnote{kubernetes-sigs/inference-perf Issue \#139 (July 2025) scopes exactly this feature; \texttt{vllm-cost-meter} can be read as a standalone realization of that RFE.} No prior work combines all four: a parameterized cost function calibrated across architectures and hardware families, a live-server instrument that reads the operator's own Prometheus metrics, and open data and tool. This paper does.

% ============================================================
% 3. FRAMEWORK
% ============================================================
\section{The Concurrency-Aware Cost Framework}\label{sec:framework}

\subsection{Problem Formulation}

We define the \textit{effective cost-per-million-tokens} $C_{\mathrm{eff}}$ as a function of four swept variables with one held implicit:

\begin{equation}\label{eq:cost}
    C_{\mathrm{eff}} = f(H, M, Q, \lambda\,;\, L)
\end{equation}

The semicolon separates the four axes we empirically sweep ($H$, $M$, $Q$, $\lambda$) from the latency-SLO parameter $L$, which we hold implicit in the sweeps reported in this paper (all configurations are measured without an SLO cutoff, and $L$ would enter as an admission-control filter on the measured distribution). Incorporating $L$ as an active sweep dimension is future work; \S\ref{sec:limitations} discusses the scope implications.

\noindent where:
\begin{itemize}[leftmargin=*,topsep=2pt,itemsep=1pt]
    \item $H$ = Hardware specification (GPU type, count, memory, interconnect)
    \item $M$ = Model architecture (parameter count, dense vs.\ MoE, attention heads)
    \item $Q$ = Quantization precision (FP16, FP8, INT8, INT4)
    \item $\lambda$ = Offered request arrival rate (requests/sec; average and peak). The number of in-flight requests resident in the engine is an emergent quantity governed by Little's Law ($N \approx \lambda \cdot W$ for mean residence time $W$), not a directly swept variable in this study
    \item $L$ = Latency SLO (target TTFT in milliseconds)
\end{itemize}

\paragraph{Offered load vs.\ in-flight concurrency.} Throughout this paper we use $\lambda$ to denote the operator-controlled \emph{offered request rate} (requests/sec arriving at the server), not the in-flight batch size of the inference engine. The two quantities are related but distinct: in-flight concurrency is what the GPU actually batches at any instant and is set by $\lambda$ together with per-request residence time $W$ via Little's Law. The methodology is ``concurrency-aware'' in the sense that it exposes the operator's offered-load operating point as a first-class input to cost---rather than assuming a fixed GPU utilization---and reports how the realized in-flight batch (and therefore $C_{\mathrm{eff}}$) responds.

\paragraph{Variables held constant.}
Five additional parameters could in principle enter~\eqref{eq:cost} but are held constant here to isolate the effect of~$\lambda$. \textbf{(i)~Inference engine}: vLLM defaults with continuous batching and PagedAttention~\cite{vllm2023}; cross-engine benchmarks~\cite{vidur2024,semianalysis2025} report that engine choice shifts~$\Theta_{\max}$ but not the shape of~$U(\lambda, L)$. \textbf{(ii)~Pricing mode}: $P_{\text{GPU}}$ is the Azure on-demand list price; reserved and spot instances multiply~$C_{\mathrm{eff}}$ by a roughly constant scalar ($0.3$--$0.7\times$) and do not alter offered-load--cost dynamics. \textbf{(iii)~Input:output token ratio}: uniform 512:256, drawn from the median of ShareGPT-derived distributions used in serving workloads~\cite{melange2024}; variable ratios affect prefill-vs-decode balance but not the underutilization penalty. \textbf{(iv)~Parallelism strategy}: tensor-parallel degree (TP=1 for dense models, TP=2 for Mixtral) is treated as part of~$H$. \textbf{(v)~Request arrival distribution}: Poisson at rate~$\lambda$. Production LLM traces follow Gamma or Weibull inter-arrival distributions with coefficient of variation exceeding~1~\cite{burstgpt2024,servegen2025}, making Poisson (CV${=}1$) a regularity assumption. Our supplementary Gamma probe (\S\ref{sec:sensitivity}) found negligible cost impact at CV${=}2$ on C4, but was limited to one configuration; bursty real-world workloads may amplify the penalty reported here rather than diminish it. We therefore interpret $C_{\mathrm{eff}}$ under these held-constant factors as a conservative \emph{lower bound} on the cost dispersion a production operator would observe.

\subsection{The Utilization Function}

The core insight is that GPU utilization $U$ is not an independent parameter to be assumed (as in prior work) but is itself a \textit{dependent variable} determined by the interaction of $H$, $M$, $Q$, $\lambda$, and $L$:

\begin{equation}\label{eq:util}
    U(\lambda, L \mid H, M, Q) = \frac{\Theta_{\text{achieved}}(\lambda, L)}{\Theta_{\max}(H, M, Q)}
\end{equation}

At low offered load, GPU compute units are underutilized---the GPU waits for memory transfers and cannot fill its execution pipeline. As offered load increases, realized in-flight concurrency improves utilization, but latency also increases due to queuing and batch contention. The latency SLO $L$ acts as a constraint that caps the achievable utilization.

The effective cost then becomes:

\begin{equation}\label{eq:ceff}
    C_{\mathrm{eff}} = \frac{P_{\text{GPU}} \times 10^6}{3600 \times \Theta_{\text{achieved}}(\lambda, L)}
\end{equation}

This differs fundamentally from the token-volume model, which assumes:

\begin{equation}\label{eq:cnaive}
    C_{\text{naive}} = \frac{P_{\text{GPU}} \times 10^6}{3600 \times \Theta_{\max}(H, M, Q)}
\end{equation}

The ratio $C_{\mathrm{eff}} / C_{\text{naive}} = \Theta_{\max} / \Theta_{\text{achieved}} = 1/U$ captures the \textbf{underutilization penalty}---the multiplicative factor by which naive estimates understate true costs.

\subsection{Architecture-Dependent Cost Behavior}

We hypothesize---and demonstrate experimentally in \Cref{sec:results}---that the shape of the utilization function $U(\lambda, L)$ differs systematically between dense and mixture-of-experts (MoE) architectures. Dense models exhibit more predictable scaling with realized in-flight batch size because all parameters are activated for every token. MoE models activate only a subset of parameters per token, leading to different memory access patterns and batch efficiency characteristics.

This means a single cost formula calibrated on one architecture will produce incorrect estimates for another, even on the same hardware. Model architecture must be a first-class variable in any cost estimation framework.

\subsection{The Crossover Analysis}

Given the utilization-adjusted cost $C_{\mathrm{eff}}$, we define the self-host vs.\ API crossover as the offered arrival-rate threshold $\lambda^*$ where:

\begin{equation}\label{eq:crossover}
    C_{\mathrm{eff}}(\lambda^*, L \mid H, M, Q) = C_{\text{API}}(\text{model\_tier})
\end{equation}

This crossover point shifts based on all five input variables. Prior work reports a single break-even (e.g., ``self-hosting is cheaper above 50M tokens/month''), but the actual crossover is a \textit{surface} in the $(\lambda, L, H)$ space, not a single point.

% ============================================================
% 4. EXPERIMENTAL SETUP
% ============================================================
\section{Experimental Setup}\label{sec:setup}

\subsection{Hardware Configurations}

All experiments are conducted on a single compute node equipped with two NVIDIA H100 NVL 96GB GPUs (Standard\_NC80adis\_H100\_v5, Azure ML). This configuration was chosen deliberately to demonstrate that the framework can be validated with modest, widely-available infrastructure---a single cloud node at \$13.96/hr on-demand---without requiring multi-node clusters or specialized hardware. Dense models (Llama 3.1 8B) and ultra-sparse MoE models (Qwen3-30B-A3B) each use a single H100 (TP=1); the sparse MoE model (Mixtral 8x7B) uses both H100s with tensor parallelism (TP=2). Single-GPU configurations (C1--C4) are costed at \$6.98/hr per GPU, which is \$13.96/hr/node amortized across the two H100s under the assumption that the second GPU is independently utilized by a co-tenant; a dedicated-node operator should double C1--C4 numbers. This denominator choice is made explicit in \S\ref{sec:results}.

\begin{table}[h]
\centering
\footnotesize
\caption{Hardware configuration.}
\label{tab:hardware}
\begin{tabular}{@{}ll@{}}
\toprule
\textbf{Component} & \textbf{Specification} \\
\midrule
GPU & 2$\times$ NVIDIA H100 NVL 96GB \\
VRAM & 96 GB HBM3 per GPU (192 GB total) \\
Interconnect & NVLink 900 GB/s \\
Host & 80 CPU cores, 640 GB RAM \\
Cloud SKU & Standard\_NC80adis\_H100\_v5 \\
Cost & \$13.96/hr node (\$6.98/GPU) \\
\bottomrule
\end{tabular}
\end{table}

\paragraph{Software environment.} All benchmarks were run on a single Azure Machine Learning H100 NVL node (NVIDIA driver 535.274.02, CUDA~12.8 runtime, Linux kernel 6.8.0-1044-azure). To avoid ABI conflicts between the two serving stacks, we maintain two isolated Python virtual environments: a \texttt{vllm-bench} environment pinned to vLLM~0.19.0 on PyTorch~2.10.0+cu128, and an \texttt{llm-bench} environment pinned to SGLang~0.5.10 on PyTorch~2.9.1+cu128. Both stacks are compiled against CUDA~12.8. The benchmark harness (commit \texttt{1c40823}) drives both engines through a unified sweep runner that enforces identical request streams, sampling parameters, and measurement windows.

\subsection{Model Selection}

We select three models representing a gradient of architectural sparsity, all well-supported by the vLLM inference engine:

\begin{table*}[t]
\centering
\caption{Models selected for benchmarking.}
\label{tab:models}
\begin{tabular}{@{}lllll@{}}
\toprule
\textbf{Model} & \textbf{Total} & \textbf{Active} & \textbf{Arch} & \textbf{Precision} \\
\midrule
Llama 3.1 8B Instruct & 8B & 8B & Dense & FP16, FP8 \\
Qwen3-30B-A3B Instruct & 30B & 3B & Ultra-sparse MoE & FP16, FP8 \\
Mixtral 8x7B Instruct & 46.7B & 12.9B & Sparse MoE & FP16, FP8 \\
\bottomrule
\end{tabular}
\end{table*}

\noindent The dense model (Llama 3.1 8B Instruct~\cite{llama31_2024}, 8B active parameters) runs on a single H100 GPU and represents the most common single-GPU deployment. Qwen3-30B-A3B-Instruct~\cite{qwen3_2025} (released July 2025) is an ultra-sparse MoE that activates only 3B of its 30B parameters per token, also fitting on a single H100. Mixtral 8x7B Instruct~\cite{mixtral_2024} activates 12.9B of its 46.7B parameters and runs on both H100 GPUs with TP=2. This three-model selection creates a sparsity gradient---dense (8B/8B), ultra-sparse (3B/30B), and moderate-sparse (12.9B/46.7B)---enabling analysis of how architectural sparsity affects the offered-load--cost relationship. These three models are selected to span architectural sparsity regimes, not to represent the frontier of model quality at submission time: newer releases shift the absolute serving economics but not the methodological point---that hardware, model architecture, quantization, offered load, and the latency SLO must enter cost estimation as first-class inputs.

\subsection{Benchmarking Protocol}

The headline numerical results in this paper come from vLLM~\cite{vllm2023}, which fixes the pricing denominator and eliminates cross-engine variance from the reported figures.\footnote{SGLang~\cite{sglang2024} runs share the same apparatus (see \S\ref{sec:setup}.1 for the two pinned venvs, \texttt{vllm-bench} and \texttt{llm-bench}) and the S1--S6 rows in the companion repository reproduce our methodology on that engine; we exclude them from headline numbers to keep the pricing denominator single-engine. TensorRT-LLM is out of scope. Cross-engine benchmarks~\cite{vidur2024,semianalysis2025} report that engine choice shifts $\Theta_{\max}$ but not the \emph{shape} of $U(\lambda, L)$, so a single engine is sufficient to characterize the underutilization-penalty structure we report.} We sweep across seven offered arrival-rate levels:

$$\lambda \in \{1, 5, 10, 25, 50, 100, 200\} \text{ requests/sec}$$

\noindent For each $(\lambda, H, M, Q)$ configuration, we measure:

\begin{itemize}[leftmargin=*,topsep=2pt,itemsep=1pt]
    \item \textbf{TTFT} (Time to First Token)---P50, P90, P99
    \item \textbf{TPS} (Tokens Per Second)---aggregate output throughput
    \item \textbf{E2E Latency}---total request completion time (P50, P90, P99)
    \item \textbf{GPU occupancy}---nvidia-smi \texttt{utilization.gpu}, the wall-clock fraction of time a kernel is resident. Reported for completeness only; it is \emph{not} the economic utilization $U{=}\Theta_{\text{achieved}}/\Theta_{\max}$ defined in \S\ref{sec:framework} and can read high even at $\lambda{=}1$ (a single resident decode kernel keeps the GPU nominally busy while it is arrival-limited at batch size~1), so we never treat it as a proxy for $U$
    \item \textbf{GPU Memory Utilization}---VRAM occupancy
\end{itemize}

Each configuration runs 100 warmup requests (discarded) followed by 500 measured requests. We use synthetic random-token prompts with fixed input length 512 tokens and output length 256 tokens, with per-run random seeds to prevent cross-run prefix cache contamination. Prefix caching is explicitly disabled on all servers (\texttt{--no-enable-prefix-caching}) to isolate offered-load effects from cache artifacts. Arrivals follow a Poisson process (\texttt{--burstiness 1.0}).

\paragraph{Deliberate Design Choices.}
Several methodological decisions deserve explicit justification, as they may draw scrutiny but are intentional.

\textit{Random synthetic tokens instead of ShareGPT or conversational traces.} We use uniform random token sequences rather than real chat datasets for three reasons: (1) ShareGPT and similar datasets contain repeated system prompts and shared prefixes that produce prefix-cache hits even with APC nominally disabled, biasing throughput upward in ways that don't generalize across deployments; (2) fixed-length synthetic workloads give exact control over the input/output length distribution, enabling perfectly reproducible results that any reader can replicate without access to proprietary data; and (3) the cost dynamics we study---how GPU utilization changes with offered request rate and the realized in-flight batch it creates---are governed by the batch-filling behavior of the scheduler, not by prompt semantics. A reviewer who argues that ``real workloads would show different numbers'' is correct: production workloads with prefix sharing and caching enabled would show \textit{lower} penalties. Our numbers are intentional upper bounds.

\textit{Prefix caching and speculative decoding disabled.} We disable automatic prefix caching (\texttt{--no-enable-prefix-caching}) to isolate pure offered-load effects on GPU utilization. Enabling APC would artificially inflate throughput for repeated prefixes, masking the underlying relationship between request rate and compute efficiency. Our goal is to measure the cost structure, not to optimize it. Note that APC savings accrue in regimes where prefill is a meaningful fraction of batch time (moderate-to-high $\lambda$ with shared prefixes); at the near-idle $\lambda{=}1$ rps extremum that drives the headline underutilization penalty, the cost is primarily GPU-rental arithmetic against a trickle of tokens, and is structurally insensitive to prefix-cache state regardless of workload.

\textit{Poisson arrivals (\texttt{--burstiness 1.0}).} This is the standard queueing-theory arrival model for open-loop load generation and is used by the majority of LLM serving benchmarks including vLLM's own benchmark harness. It models independent, memoryless arrivals with no request batching at the client side; burstier-than-Poisson production arrivals are treated empirically in \S\ref{sec:sensitivity}.

These design choices---random tokens, no prefix caching, no speculative decoding, no chunked-prefill tuning---deliberately establish a \textit{conservative performance floor}. Production workloads with repeated system prompts, shared prefixes, or optimized inference pipelines would achieve higher throughput and therefore lower cost-per-token than reported here. All cost figures in this paper should be interpreted as upper bounds on achievable cost; real-world deployments with standard optimizations enabled would only improve on these numbers.

\textbf{Configuration matrix:} 1 hardware node $\times$ 3 models $\times$ 2 precisions = 6 server configurations. Each swept at 7 arrival-rate levels produces \textbf{42 benchmark runs} (approximately 4.5 hours of total GPU time, under \$65 at on-demand pricing). All code, data, and a reusable cost estimation tool are released as open-source under an MIT license; see \S\ref{sec:artifact}.

\paragraph{Raw saturation vs.\ goodput.} We report $\Theta_{\max}$ as the sustained tokens-per-second achievable at peak offered load under a fixed I/O shape with no latency bound imposed. In the goodput literature~\cite{inferenceperf2025,semianalysis2025}, this is the \emph{raw saturation throughput}, distinct from \emph{goodput}, which would additionally constrain on a TTFT/TPOT/E2EL SLO. Raw saturation is an upper bound on goodput: any meaningful production SLO will reduce the achievable arrival rate, and therefore the effective tokens per dollar. We intentionally measure raw saturation to isolate the engine-plus-hardware ceiling from SLO-policy choices; goodput-bounded extensions are a natural direction for community contribution, and our companion tool's \texttt{workload\_protocol} schema is designed to accommodate crowd-sourced curves with self-declared SLO bounds.

\subsection{Derived Metrics}

From raw measurements, we compute:

\begin{itemize}[leftmargin=*,topsep=2pt,itemsep=1pt]
    \item \textbf{Effective cost-per-million-tokens:} $C_{\mathrm{eff}} = \frac{P_{\text{GPU}} \times 10^6}{\text{TPS} \times 3600}$ \,(\$/M \emph{output} tokens). Throughout, TPS is aggregate \emph{output}-token throughput, so every $C_{\mathrm{eff}}$ figure is dollars per million \emph{output} tokens; this matches the output-token basis of the API list prices we compare against (\S\ref{sec:crossover}). Input (prefill) tokens consume compute but are not in the denominator.
    \item \textbf{Utilization rate:} $U{=}\Theta_{\text{achieved}}(\lambda)/\Theta_{\max}$, the throughput ratio defined in \S\ref{sec:framework} (distinct from the nvidia-smi GPU-occupancy metric)
    \item \textbf{Underutilization penalty:} $C_{\mathrm{eff}} / C_{\text{naive}}$ at each operating point
    \item \textbf{Crossover arrival rate:} the $\lambda$ where $C_{\mathrm{eff}}$ equals API pricing for GPT-5.5, Gemini 3.1 Pro, and Claude Sonnet 4.6
\end{itemize}

% ============================================================
% 5. RESULTS
% ============================================================
\section{Results}\label{sec:results}

We report results across all 42 benchmark configurations. GPU cost follows the denominator convention introduced in \S\ref{sec:setup} (and Table~\ref{tab:hardware}): \$6.98/hr per GPU for single-GPU models (C1--C4, assuming co-tenant utilization of the second H100) and \$13.96/hr for Mixtral (C5--C6, TP=2, both GPUs occupied). The \emph{shape} of $C_{\mathrm{eff}}(\lambda)$ is unchanged under the alternative dedicated-node denominator; only the absolute level of C1--C4 doubles.

\subsection{Offered Load vs.\ Effective Cost}

\begin{figure}[t]
\centering
\includegraphics[width=\columnwidth]{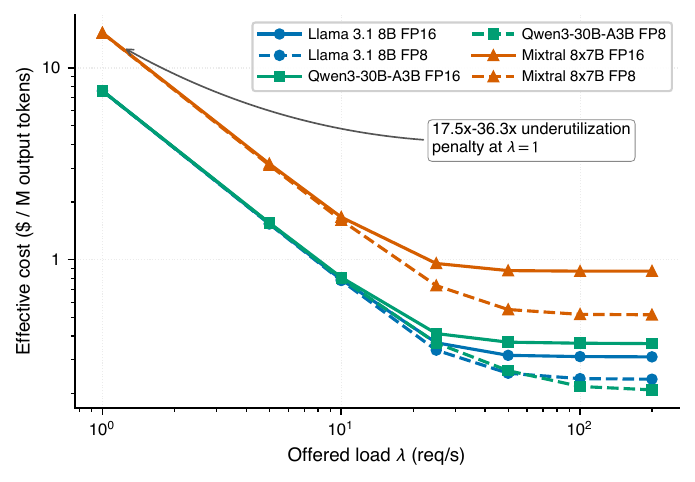}
\caption{$C_{\mathrm{eff}}$ vs.\ offered load $\lambda$ across six H100 configurations (log-log); solid lines are FP16, dashed FP8. Effective cost drops by roughly $20\times$ over the first 25~rps, then flattens as the engine saturates. Calculators that omit $\lambda$ and assume a fixed utilization or peak throughput therefore misreport cost by close to an order of magnitude in the low-$\lambda$ regime. At $\lambda \geq 50$ the server is queue-limited; throughput and cost use completed-request statistics within the measurement window. Note the architecture signal: at $\lambda{=}1$ the two single-GPU FP16 models (Llama, Qwen) sit at an identical $\approx$\$7.60/MTok---idle throughput is arrival-limited ($\approx$255~tok/s) and hence model-independent---so Mixtral's $\sim2\times$-higher idle cost reflects its two-GPU (\$13.96/hr) denominator, not its parameter count.}
\label{fig:cost_concurrency}
\end{figure}

\Cref{fig:cost_concurrency} shows $C_{\mathrm{eff}}$ across all six configurations. At $\lambda{=}1$ rps, effective costs range from \$7.60/MTok (Llama FP16/FP8) to \$15.25/MTok (Mixtral FP16)---the near-idle GPU cost amortized over a trickle of tokens. As offered load increases, $C_{\mathrm{eff}}$ falls precipitously: by $\lambda{=}25$ rps all single-H100 models are within 20\% of peak-throughput cost.

The naive cost model underestimates self-hosted costs by \textbf{17.5--36.3$\times$}. That is the $1/U$ gap between a utilization-naive estimate (any calculator that assumes the GPU runs at its $\Theta_{\max}$) and what you actually pay when the engine is arrival-limited at 1~rps. Prior informal estimates put the error at 3--10$\times$; ours shows they were themselves off by another 3--10$\times$. Even at 10~rps---an arrival rate most practitioners would not call ``low''---the naive model is still wrong by 2.5$\times$ on dense Llama and 3.8$\times$ on ultra-sparse Qwen. The naive model is not merely optimistic; it omits the dominant cost term.

\subsection{Architecture-Dependent Scaling}

The three FP16 solid lines in \Cref{fig:cost_concurrency} isolate the architectural families. At peak throughput, Llama 3.1 8B achieves 6{,}238~tok/s (\$0.311/MTok), Qwen3-30B-A3B achieves 5{,}319~tok/s (\$0.364/MTok), and Mixtral 8x7B on TP=2 achieves 4{,}454~tok/s at \$0.871/MTok; the three curves converge to within $3\times$ at saturation.

Qwen's lower throughput than Llama---despite activating only 3B parameters---reflects two structural costs: (1) MoE routing dispatch overhead per token, and (2) the full 30B parameter set occupying VRAM, constraining memory bandwidth. The \textit{shape} of the scaling curve is, however, comparable between the two: the ultra-sparse Qwen model's cost cliff from $\lambda{=}1$ to $\lambda{=}10$ rps (9.4$\times$ drop) is marginally shallower than dense Llama's (9.7$\times$ drop), indicating that routing overhead does not meaningfully shift the per-request amortization curve at this scale. The cost-structure differences between dense and ultra-sparse MoE therefore manifest primarily as level shifts (higher idle cost, different saturation floor) rather than slope shifts, but are large enough in magnitude that a single cost formula calibrated on dense models will systematically misestimate MoE costs.

This active-parameter dominance holds in the batched decode regime measured here ($\lambda \geq 1$ rps, batch sizes reaching hundreds), where per-token arithmetic scales with active parameters while weight-memory movement is amortized across the batch. At small batch sizes or in prefill-dominated regimes, total parameters drive memory bandwidth cost and the active-vs-total distinction narrows~\cite{mixtral_2024}---a caveat that applies to the $\lambda{=}1$ edge of our sweep and to any deployment with long prompts and short outputs.

\subsection{Quantization Impact}

\begin{figure}[t]
\centering
\includegraphics[width=\columnwidth]{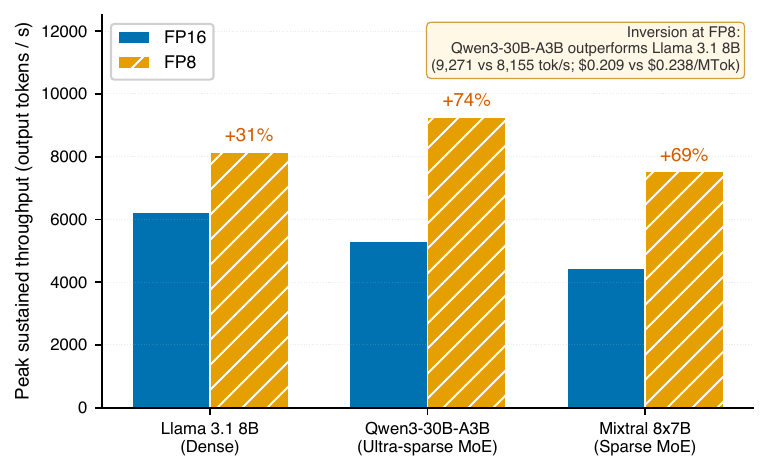}
\caption{FP8 gives the dense model +31\% peak throughput and the two MoE models +69\% (Mixtral) and +74\% (Qwen)---a roughly $2.2$--$2.4\times$ larger win for the MoE architectures. Quantization behaves as an MoE-first optimization that also helps dense models.}
\label{fig:quant}
\end{figure}

FP8 quantization reduces $C_{\mathrm{eff}}$ at every arrival-rate level (\Cref{fig:quant}). Peak throughput gains are: \textbf{+31\%} for Llama 3.1 8B (6{,}238$\to$8{,}155~tok/s), \textbf{+74\%} for Qwen3-30B-A3B (5{,}319$\to$9{,}271~tok/s), and \textbf{+69\%} for Mixtral 8x7B (4{,}454$\to$7{,}524~tok/s). The larger FP8 gain for MoE models is consistent with their memory-bandwidth-bound profile: sparse activation requires many small, scattered parameter reads, and halving parameter precision directly doubles effective bandwidth.

A notable emergent result: at peak offered load, \textbf{Qwen3-30B-A3B FP8 (\$0.209/MTok) is cheaper than Llama 3.1 8B FP8 (\$0.238/MTok)} despite having 30B total parameters. On this dense/ultra-sparse pair, the ordering inverts the common intuition that smaller models are cheaper to serve: \textit{active} parameter count and memory-access pattern, rather than total model size, appear to govern the saturation cost. We treat this as suggestive---it rests on a single architecture pair with a $\sim$12\% margin---and defer the larger-dense (e.g.\ Llama~3.1~70B) plus third-sparsity-ratio validation that would license a general claim to \S\ref{sec:conclusion}.

\subsection{The Underutilization Penalty}

\begin{table}[h]
\centering
\scriptsize
\setlength{\tabcolsep}{3.5pt}
\caption{Underutilization penalty and latency tails for the dense reference model (Llama 3.1 8B FP16, single H100, \$6.98/hr). $C_{\text{naive}} = \$0.311$/MTok at peak throughput. TTFT and TPOT percentiles are measured (vLLM \texttt{benchmark\_serving}); at $\lambda \geq 50$ the server is queue-limited, so throughput, cost, and latency use completed-request statistics within the measurement window. At the $\lambda{=}50$ cost floor both tails exceed the example SLO of \S\ref{sec:serverless_fallacy} (TTFT$_{p99}\leq300$\,ms, TPOT$_{p99}\leq50$\,ms).}
\label{tab:penalty}
\begin{tabular}{@{}rrrrrr@{}}
\toprule
$\lambda$ & TTFT$_{p50}$ & TTFT$_{p99}$ & TPOT$_{p99}$ & $C_{\mathrm{eff}}$ & Penalty \\
(rps) & (ms) & (ms) & (ms) & (\$/MTok) & \\
\midrule
1   & 37.9    & 54.3    & 7.0  & 7.60 & 24.4$\times$ \\
5   & 44.1    & 713.3   & 16.4 & 1.54 & 5.0$\times$  \\
10  & 49.4    & 84.5    & 12.3 & 0.79 & 2.5$\times$  \\
25  & 113.5   & 259.0   & 37.8 & 0.37 & 1.18$\times$ \\
50  & 417.9   & 994.8   & 71.3 & 0.32 & 1.02$\times$ \\
100 & 2{,}561 & 5{,}011 & 76.1 & 0.31 & 1.00$\times$ \\
200 & 3{,}748 & 7{,}437 & 76.8 & 0.31 & 1.00$\times$ \\
\bottomrule
\end{tabular}
\end{table}

\Cref{tab:penalty} quantifies the penalty for our dense reference model. At $\lambda{=}1$ rps, effective cost is 24.4$\times$ the naive estimate. The penalty declines non-linearly: by $\lambda{=}25$ rps it falls to 1.18$\times$, and beyond $\lambda{=}50$ rps throughput approaches the hardware ceiling. Importantly, at $\lambda \geq 100$ rps, throughput plateaus at $\sim$6{,}230~tok/s while TTFT P50 rises to 2{,}561--3{,}748~ms, indicating queue saturation. This defines the model's \textit{saturation threshold}---the arrival rate above which additional load degrades latency without improving cost. The latency tails confirm that this cost floor is a \emph{no-SLO} floor: at the $\lambda{=}50$ minimum-cost point TTFT$_{p99}$ has already reached 994.8~ms and TPOT$_{p99}$ 71.3~ms---both well beyond the example production contract of \S\ref{sec:serverless_fallacy} (TTFT$_{p99}\leq300$~ms, TPOT$_{p99}\leq50$~ms)---and the C2 repeat campaign shows TTFT$_{p50}$ run-to-run CV rising to $8.73\%$ at the same load (\S\ref{sec:stability}). An operator bound by such an SLO must back off to a lower arrival rate, where $C_{\mathrm{eff}}$ is strictly higher; the raw-saturation $C_{\min}$ reported here is therefore a \emph{lower} bound on goodput-constrained cost---equivalently, an upper bound on the tokens-per-dollar an SLO-bound operator can achieve---exactly as the goodput-vs-saturation distinction in \S\ref{sec:setup} and the operational workflow in \S\ref{sec:operational_framework} prescribe.\footnote{The $\lambda{=}5$ row's tails (TTFT$_{p99}$ 713.3~ms, TPOT$_{p99}$ 16.4~ms) are inflated by a handful of outlier requests on this single, unrepeated run, with the bulk of the distribution unaffected (TTFT P50/P90 of 44.1/56.8~ms); monotone tail growth with load resumes by $\lambda{=}10$ (TTFT$_{p99}$ 84.5~ms, TPOT$_{p99}$ 12.3~ms). The P50, throughput, and cost columns are unaffected.}

\begin{figure}[t]
\centering
\includegraphics[width=\columnwidth]{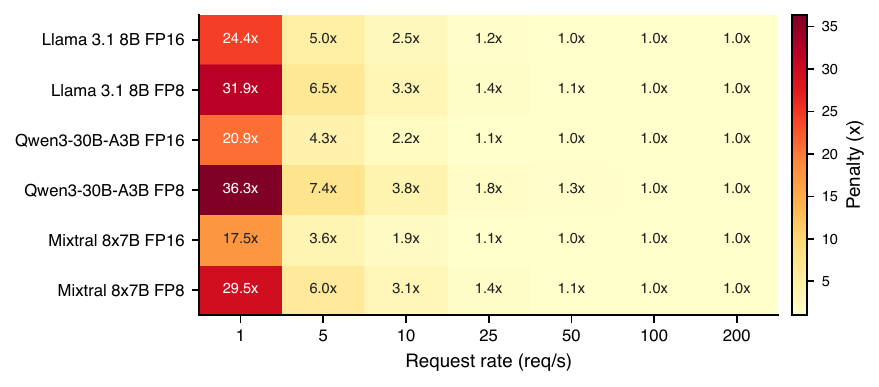}
\caption{Underutilization penalty (idle-edge $C_{\mathrm{eff}}$ over saturation $C_{\mathrm{sat}}$) by config and offered load. The penalty lives almost entirely in the $\lambda{=}1$ column (17.5--36.3$\times$) and collapses to $\approx$1.0$\times$ by $\lambda{=}50$ across every config. Within each model the FP8 variant carries the larger penalty---quantization lowers the saturation floor more than the idle-edge cost---peaking at the darkest cell, 36.3$\times$ for Qwen3-30B-A3B FP8.}
\label{fig:heatmap}
\end{figure}

\Cref{fig:heatmap} extends this to all configurations. The maximum penalty is \textbf{36.3$\times$} for Qwen3-30B-A3B FP8 at $\lambda{=}1$ rps. Penalties at $\lambda{=}1$ rps range from 17.5$\times$ (Mixtral FP16) to 36.3$\times$ (Qwen FP8), converging to $\approx$1.0$\times$ by $\lambda{=}50$ rps across all configurations. The consistent pattern across architectures and precisions confirms that the underutilization trap is a structural property of discrete GPU rental cost, not an artifact of any specific model.

\subsection{SLO-Conditioned Operating Points}\label{sec:slo}

The cost floors quoted so far are \emph{unconstrained} saturation values: they are the cheapest $C_{\mathrm{eff}}$ a config reaches at any $\lambda$. In production they are typically unreachable, because the latency-$L$ argument of $C_{\mathrm{eff}}=f(H,M,Q,\lambda,L)$ is not a free variable---it is a \emph{commitment}. An operator who signs a service-level agreement (a B2B latency SLA, an internal interactive-product target) fixes $L$, and a fixed $L$ caps the offered load the deployment may sustain before its tail breaches, which in turn pins the operating point on the $C_{\mathrm{eff}}(\lambda)$ curve. The SLA is therefore a price: tightening it forces a lower $\lambda$ and a higher cost per token.

\begin{figure}[t]
\centering
\includegraphics[width=\columnwidth]{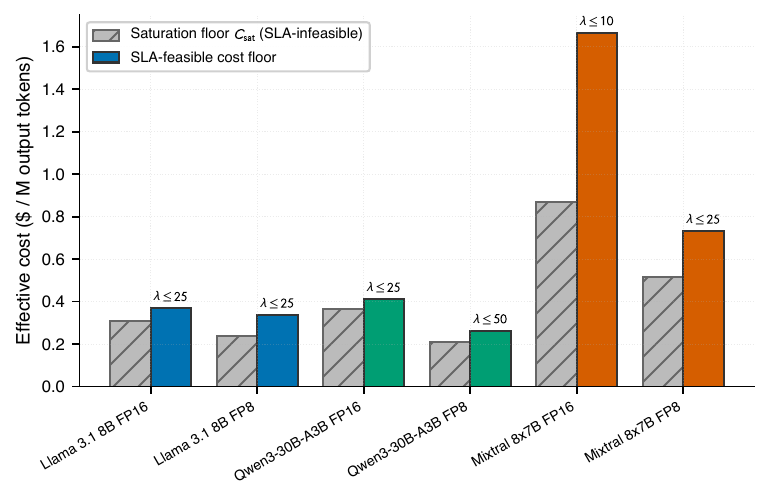}
\caption{SLA-feasible cost floor (solid, per-config color) vs.\ the unconstrained saturation floor $C_{\mathrm{sat}}$ (grey, hatched) under a single fixed SLA: TTFT p99 $\leq 300$\,ms and TPOT p99 $\leq 50$\,ms. Annotations give the highest SLA-feasible offered load. $C_{\mathrm{sat}}$ is unreachable under this SLA---it lands at $\lambda{=}100$--$200$ where TTFT p99 runs to multiple seconds (\Cref{tab:slo_operating})---so the SLA-feasible floor, not $C_{\mathrm{sat}}$, is the cost an SLA-bound operator actually pays.}
\label{fig:slo}
\end{figure}

To make this concrete on the existing corpus we evaluate one fixed SLA used as the running example in \S\ref{sec:serverless_fallacy}: TTFT p99 $\leq 300$\,ms \emph{and} TPOT p99 $\leq 50$\,ms. For each config we take the highest $\lambda$ in the 7-point sweep that satisfies both bounds and report its $C_{\mathrm{eff}}$ (\Cref{tab:slo_operating}, \Cref{fig:slo}).

\begin{table}[t]
\centering
\footnotesize
\setlength{\tabcolsep}{2.5pt}
\caption{Cost of honoring a fixed SLA (TTFT p99 $\leq 300$\,ms, TPOT p99 $\leq 50$\,ms). $\lambda_{\max}$ is the highest SLA-feasible offered load; the premium is the SLA-feasible cost over the unconstrained saturation floor $C_{\mathrm{sat}}$. Every $C_{\mathrm{sat}}$ here lives at $\lambda{=}100$--$200$ with a multi-second TTFT p99 (2.2--15.7\,s), i.e.\ SLA-infeasible.}
\label{tab:slo_operating}
\begin{tabular}{lrrrr}
\toprule
Config & SLA $\lambda_{\max}$ & \$/MTok & $C_{\mathrm{sat}}$ & premium \\
       & (rps)               & at SLA  &                    & \\
\midrule
Llama 8B FP16   & 25 & 0.368 & 0.311 & 1.18$\times$ \\
Llama 8B FP8    & 25 & 0.337 & 0.238 & 1.42$\times$ \\
Qwen3-A3B FP16  & 25 & 0.411 & 0.364 & 1.13$\times$ \\
Qwen3-A3B FP8   & 50 & 0.263 & 0.209 & 1.26$\times$ \\
Mixtral FP16    & 10 & 1.666 & 0.871 & \textbf{1.91$\times$} \\
Mixtral FP8     & 25 & 0.732 & 0.515 & 1.42$\times$ \\
\bottomrule
\end{tabular}
\end{table}

Two readings follow. First, the advertised floor is a fiction for any latency-bound deployment: Llama~8B~FP16's \$0.311/MTok floor lives at $\lambda{=}200$, where TTFT p99 is \textbf{7.4 seconds}---no interactive product ships that. Its cheapest \emph{shippable} cost under the SLA is \$0.368/MTok, at $\lambda{=}25$. Second, the SLA premium over the (already-unreachable) floor ranges from 1.13$\times$ to \textbf{1.91$\times$}, largest on Mixtral~8x7B~FP16, whose strict per-GPU memory budget forces the SLA-feasible load down to $\lambda{=}10$. This is a deliberately conservative slice of $L$'s effect: it compares two points on the same curve. Loosening the SLA toward batch-grade latency walks the operating point rightward toward $C_{\mathrm{sat}}$, and tightening it past our interactive example walks it leftward into the steep part of \Cref{fig:cost_concurrency}, where the penalty is the 17.5--36.3$\times$ already reported. The practical statement is that the SLA an operator commits to in a contract is, through $\lambda$, a directly priced quantity---and one no token-volume calculator can express.

\subsection{Corrected Crossover Analysis}\label{sec:crossover}

\begin{figure}[t]
\centering
\includegraphics[width=\columnwidth]{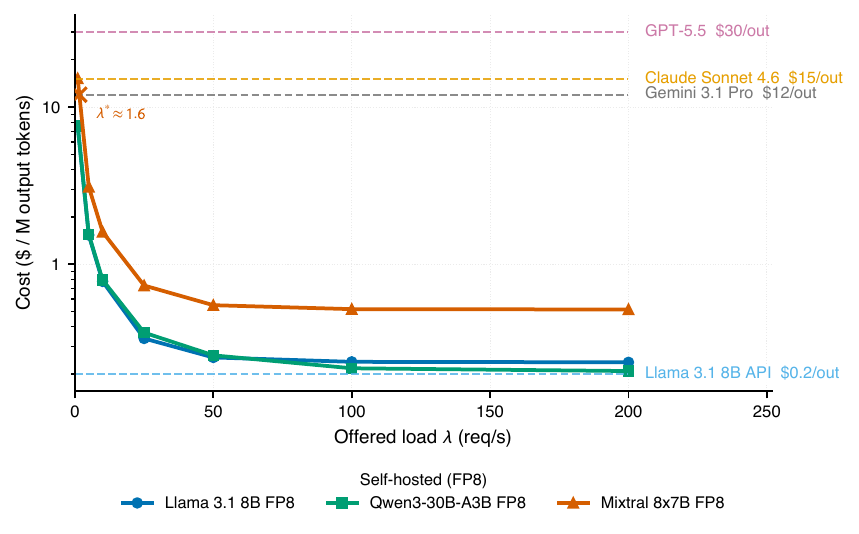}
\caption{Both axes use \emph{cost per million output tokens}. Self-hosted $C_{\mathrm{eff}}$ is GPU $\$/\mathrm{hr}$ divided by output throughput; dashed API lines are each provider's \emph{list output-token} price only (the ``/out'' suffix in the inline labels is a reminder of this). Input-token billing, prompt-caching, and batch-API discounts are \emph{not} applied to the API lines---all three would shift them down (see \S\ref{sec:crossover} caveat). Plotted self-hosted lines are the FP8 variants (the FP16 curves sit within $<1\%$ of FP8 at $\lambda{=}1$, so the crossover visual is unchanged); prose in \S\ref{sec:crossover} quotes the FP16 value ($\$15.25$/MTok) for consistency with the H100 headline sweep. The $\lambda^{\star}{\approx}1.6$ annotation on the Mixtral~FP8 curve marks its log-y-interpolated crossover against the Gemini~3.1~Pro output tier ($\$12$/MTok), the lowest-priced non-open-source reference line in the panel; the Claude~Sonnet~4.6 tier at $\$15$/MTok is crossed at $\lambda{\approx}1$, while the GPT-5.5 tier ($\$30$/MTok) sits above the Mixtral curve at every measured $\lambda$. Under naive utilization math every self-hosted line sits below every API line; under real utilization math Mixtral at 1~rps is more expensive than Claude~Sonnet~4.6's output tier. The crossover is not a point; it is a threshold that moves with traffic and with the reference tier. API prices are public list prices as of 2026-06-09~\cite{openai_pricing2025,anthropic_pricing2025,google_pricing2026}, included only as external economic reference tiers; the benchmark corpus itself is fixed for reproducibility.}
\label{fig:crossover}
\end{figure}

\Cref{fig:crossover} plots $C_{\mathrm{eff}}(\lambda)$ against commercial API reference prices. Under the naive model, all configurations appear permanently cheaper than API alternatives---Llama FP8 at \$0.238/MTok is over 50$\times$ cheaper than Gemini~3.1~Pro (\$12.00/M output tokens~\cite{google_pricing2026}). Under the utilization-adjusted model, the picture differs.

At $\lambda{=}1$ rps, Mixtral FP16 costs \$15.25/MTok---more expensive than Claude Sonnet 4.6 (\$15.00/M output tokens). At $\lambda{=}5$ rps it falls to \$3.16/MTok, below all API tiers. The naive analysis concludes Mixtral self-hosting is economical at any traffic level; the adjusted analysis reveals a \textit{crossover threshold near 1.5--2 rps} below which the lower-priced reference tiers (Gemini~3.1~Pro, Claude~Sonnet~4.6) are actually cheaper.

For Llama and Qwen models, the lower single-GPU cost means the crossover is below 1 rps---self-hosting is cost-effective even at minimal traffic, if the GPU is already provisioned---\emph{but only under the \$6.98/hr co-tenant denominator of \S\ref{sec:setup}}. Under the dedicated-node \$13.96/hr rate the entire single-GPU $C_{\mathrm{eff}}(\lambda)$ curve doubles, roughly doubling the low-$\lambda$ cost (e.g.\ Llama/Qwen idle rises from $\approx$\$7.60 to $\approx$\$15.20/MTok) and pushing the API crossover to a higher arrival rate. Operators on a dedicated node should read the C1--C4 numbers at the doubled rate. (The $\lambda{<}1$ region is extrapolated outside our 7-point sweep; the exact crossover location below 1 rps is a modeled continuation of the measured $C_{\mathrm{eff}}(\lambda)$ curve, not a directly observed operating point.) However, the effective cost at 1 rps (\$7.60/MTok) is 24$\times$ the naive estimate, compressing the apparent cost advantage over the lowest-priced reference tier (Gemini~3.1~Pro, \$12.00/M output) from 98\% (naive) to roughly a third at $\lambda{=}1$ rps.\footnote{At the specific $\lambda{=}1$ operating point: $1 - 7.60/12.00 \approx 37\%$. The exact figure moves with the tier's list price, which changes on vendor schedule---and under the dedicated-node denominator ($\approx$\$15.20/MTok idle) the advantage inverts against the \$12 tier and vanishes against the \$15 tier. The qualitative point---expected savings compress sharply at low-traffic operating points---is what matters for infrastructure planning.} An organization planning infrastructure on the naive model will be surprised to find that at realistic low-traffic operating points, most of the expected advantage evaporates.

\paragraph{Caveat on the API comparison.} \Cref{fig:crossover} compares self-hosted $C_{\mathrm{eff}}$ against \emph{list output-token} prices for the commercial APIs, to keep the comparison reproducible from public pricing pages. Three real-world discounts are not applied and would move the API lines down: (i)~list \emph{input}-token prices are typically $\sim$5$\times$ lower than output prices, so a workload whose output:input ratio is smaller than our 512:256 assumption has a lower blended API cost; (ii)~prompt-caching tiers now offered by Anthropic, OpenAI, and Google can reduce cached-input token cost by roughly an order of magnitude for workloads with substantial prefix sharing (system prompts, RAG preambles, agentic loops); (iii)~batch APIs ($\sim$50\% off) and volume/enterprise contracts further compress the effective API price. In all three cases the comparison in \Cref{fig:crossover} is conservative \emph{in favour of} self-hosting---applying these discounts would push the crossover threshold to \emph{higher} $\lambda$, not lower.

\subsection{Sensitivity to Held-Constant Factors}\label{sec:sensitivity}

\S\ref{sec:framework} enumerated five factors held constant in the main sweep: engine, pricing mode, input:output ratio, parallelism, and arrival distribution. Two of these---the I/O ratio and the arrival distribution---admit direct empirical probes on the same hardware, as does a third held-constant protocol choice: prefix caching (disabled in the main sweep). We ran a targeted supplementary sweep across these three factors to quantify how far the headline underutilization penalty generalizes when those knobs are unfrozen. We report the results here rather than relegating them to an appendix because they materially bound the scope of our headline claim.

\begin{figure}[t]
\centering
\includegraphics[width=\columnwidth]{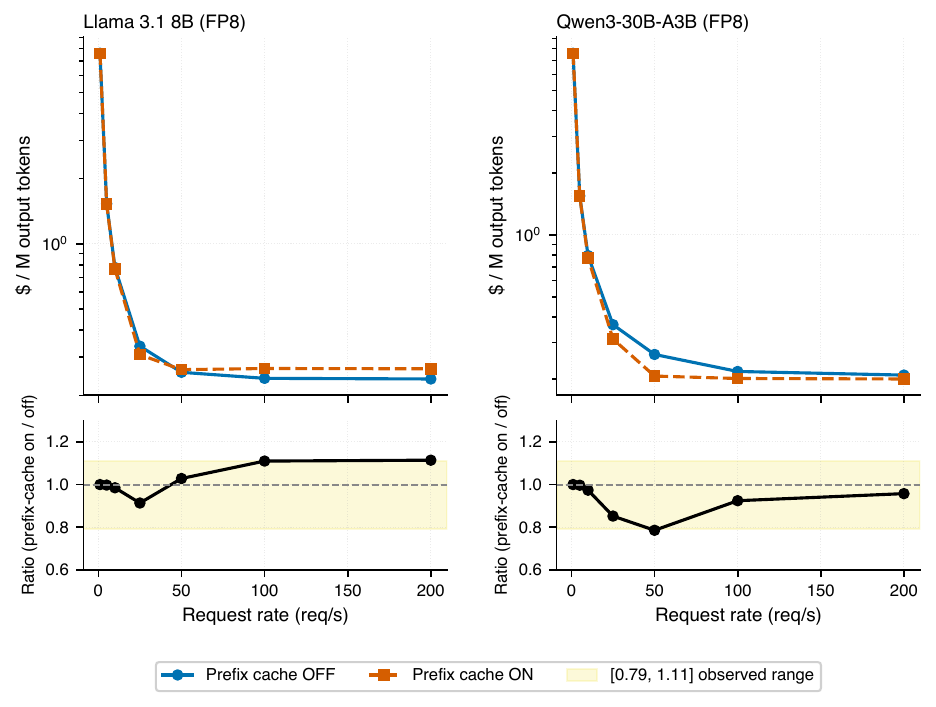}
\caption{Prefix-cache OFF vs.\ ON on random-token workloads (no shared prefixes). APC is inert by construction; the small $+11\%$ overhead visible at $\lambda{\in}\{100, 200\}$ on Llama C2 is APC's hashing bookkeeping, not a serving pathology. Production workloads with real prefix sharing go the OTHER way. The main-sweep numbers are not artifacts of prefix-cache state.}
\label{fig:prefix_cache}
\end{figure}

\paragraph{Prefix caching (\Cref{fig:prefix_cache}).} On our random-token workload no two prompts share a prefix, so vLLM's automatic prefix cache (APC)~\cite{vllm2023} cannot hit and therefore cannot save any prefill compute---its hit-rate is $\sim$0 by construction. What the data actually show is the \emph{cost} of running APC's hashing and block-bookkeeping path with no corresponding savings: across all offered-load levels for C2 (Llama~3.1~8B FP8) and C4 (Qwen3-30B-A3B FP8), the pc-on/pc-off cost ratio lies in $[0.79, 1.11]$ with a median near $1.00$, and in the worst tail---C2 at $\lambda \in \{100, 200\}$---APC-on is up to $+11\%$ \emph{more expensive} than APC-off. That is the correct direction: on prefix-free inputs, turning APC on adds overhead without adding any cache hits. The direction \emph{reverses} on production workloads with real prefix sharing---shared system prompts, agentic loops, RAG preambles---where APC, KV-cache disaggregation across tiers (Mooncake~\cite{mooncake2024}), and KV-cache reuse systems (LMCache~\cite{lmcache2025}) deliver substantial savings; our random-token protocol deliberately removes that effect to isolate the offered-load cost structure. The point we need for this paper is unchanged: the main-sweep $C_{\mathrm{eff}}(\lambda)$ numbers are not artifacts of APC state, and the conservative-lower-bound framing in \S\ref{sec:setup} is supported by data.

\begin{figure}[t]
\centering
\includegraphics[width=\columnwidth]{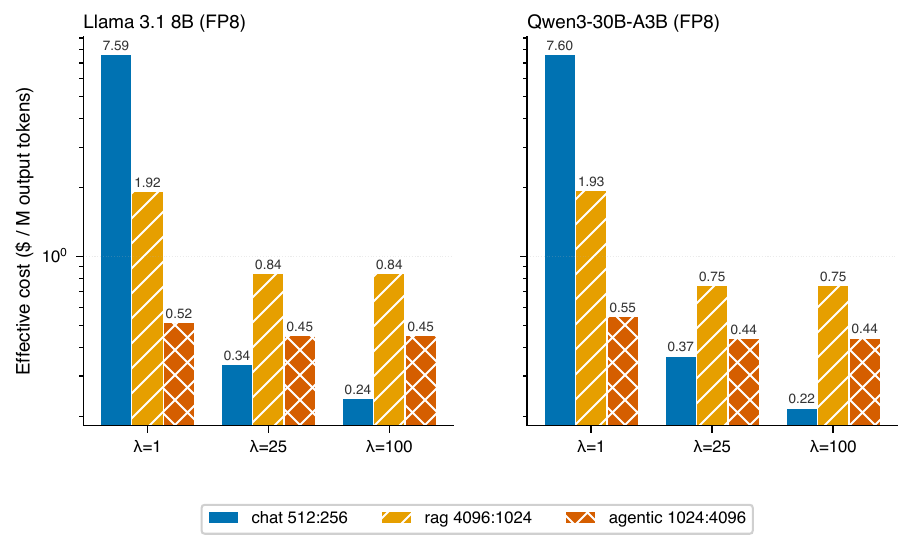}
\caption{Effective cost at $\lambda{\in}\{1, 25, 100\}$ across three I/O shapes (chat~512:256, RAG~4096:1024, agentic~1024:4096) for C2 and C4 with prefix caching off and Poisson arrivals. The RAG/chat cost ratio is non-monotonic: RAG is \emph{cheaper} than chat at $\lambda{=}1$ (long prefill amortises cold-start cost), 2.4--2.7$\times$ at $\lambda{=}25$, and peaks at 3.2--3.7$\times$ at $\lambda{=}100$; prompt-heavy workloads dominate cost at high load while preserving the offered-load sensitivity structure of the framework.}
\label{fig:io_shapes}
\end{figure}

\paragraph{I/O shape (\Cref{fig:io_shapes}).} Holding the arrival process Poisson and APC off, we replaced the 512:256 chat shape with a RAG shape (4096:1024) and an agentic shape (1024:4096) at $\lambda{\in}\{1, 25, 100\}$. The RAG/chat cost ratio is non-monotonic in~$\lambda$: at $\lambda{=}1$ RAG is actually \emph{cheaper} than chat (the long 4K-token prefill raises effective tokens-per-second at a near-idle GPU, amortising the fixed cold-start cost over more tokens and lowering \$/MTok). At the mid-load $\lambda{=}25$ RAG is roughly 2.4--2.7$\times$ the chat baseline (C4 Qwen 2.39$\times$, C2 Llama 2.73$\times$), and at the saturation-adjacent $\lambda{=}100$ it peaks at 3.2--3.7$\times$ while TTFT p50 is pushed into the hundreds of seconds (queued prefill on 4K-token prompts). The agentic shape---dominated by output tokens---lands between chat and RAG throughout. Crucially, the \emph{shape} of the $C_{\mathrm{eff}}(\lambda)$ curve is preserved across I/O mixes: every shape still exhibits the concave fall-off from cold-start cost to saturation that motivates the framework. What changes is the y-axis offset, not the qualitative penalty structure. The underutilization story is robust to I/O mix; the absolute cost at a given $\lambda$ is not.

\paragraph{Arrival burstiness.} Production LLM traces exhibit bursty, non-Poisson arrivals~\cite{burstgpt2024,servegen2025}. On C4 chat with APC off, we reran $\lambda{\in}\{10, 50, 100\}$ under Gamma inter-arrivals at CV$=$2 (\texttt{--burstiness 0.25}) against a \emph{matched} Poisson (CV$=$1) baseline---same harness version, same 120\,s time-based measurement window, same prefix-cache and I/O-shape settings, varying only the inter-arrival distribution. The effect is negligible at all three arrival rates: Gamma-CV$=$2 is $0.3\%$ cheaper at $\lambda{=}10$ ($0.997\times$ ratio), $0.5\%$ cheaper at $\lambda{=}50$ ($0.995\times$), and $0.8\%$ \emph{more expensive} at $\lambda{=}100$ ($1.008\times$). All three deltas sit well inside run-to-run noise observed across repeat measurements of the same configuration. The practical reading is that vLLM's continuous-batching scheduler absorbs a factor-of-two increase in arrival-process variance without a measurable cost signature at these arrival rates: the KV-cache and the scheduler's batch-formation window are large enough, relative to the mean inter-arrival gap, that short bursts do not meaningfully displace the steady-state packing a Poisson stream would achieve. A production workload whose arrival-process CV deviates from~$1.0$ by this much therefore neither amplifies nor compresses the headline underutilization penalty in any material way; we preserve our conservative-lower-bound characterization of $C_{\mathrm{eff}}$ under Poisson assumptions. \emph{Scope:} the Gamma probe is run on C4 only; the burstiness-invariance claim is therefore bounded to this configuration. A spot-check on ultra-sparse MoE at TP${>}1$ (where routing dispatch interacts with the scheduler differently) is deferred to future work.

% Figure 8 (Gamma vs Poisson) removed: three data points all within run-to-run
% noise (ratios 0.997/0.995/1.008) do not warrant a figure; data reported in prose.

\paragraph{Variable-length arrivals.} The three probes above use fixed-length I/O shapes; a natural concern is whether the offered-load cost cliff survives when prompt lengths \emph{vary within} a single run, as they do in production traffic. We generated a synthetic variable-length dataset with log-normal distributed input and output lengths (input median $\sim$400 tokens, p10/p90 $= 120/906$; output median $\sim$200 tokens, p10/p90 $= 68/408$) and swept $\lambda \in \{1, 10, 50, 100\}$ on C2 and C4. The result is unambiguous: the load-driven spread under variable-length arrivals is \emph{wider} than under fixed 512:256---$39.3\times$ on C2 (vs.\ $31.9\times$ fixed) and $47.6\times$ on C4 (vs.\ $36.3\times$ fixed). At $\lambda{=}1$, variable-length $C_{\mathrm{eff}}$ matches the fixed-length baseline to within $0.1\%$ (\$7.59/MTok); at saturation ($\lambda{=}100$) the variable-length workload achieves higher throughput because the scheduler can interleave short and long requests more efficiently. The cliff is steeper, not shallower, under realistic mixed-length traffic. This rules out the concern that fixed-length workloads artificially inflate the headline spread.

\paragraph{Prefix caching with real cache hits.} The prefix-cache probe in the preceding paragraph tested APC on random tokens where it cannot hit by construction. To test whether APC changes the \emph{shape} of the $C_{\mathrm{eff}}(\lambda)$ curve when it actually helps, we ran a shared-prefix workload (10 distinct 200-token prefixes, each reused across $\sim$50 prompts, 312-token suffixes, 256-token output) on C2 and C4 at $\lambda \in \{1, 10, 50, 100\}$ with APC on and off. At low offered load ($\lambda \leq 10$) APC has negligible effect (ON/OFF ratio $\approx 1.00$): the cache hit rate is high but throughput is memory-latency-bound regardless. At saturation ($\lambda = 100$) APC reduces $C_{\mathrm{eff}}$ by $20$--$22\%$ (ON/OFF ratio $0.78$--$0.80$): the saved prefill compute lets the scheduler pack more decode tokens into each batch. Crucially, both APC-on and APC-off exhibit the same characteristic cliff from $\lambda{=}1$ to $\lambda{=}50$, with spreads of $30$--$47\times$ in both cases. APC shifts the cost \emph{level} downward at high $\lambda$ but does not change the \emph{shape} of the offered-load cost curve. The underutilization story is robust to prefix caching; the absolute cost at saturation is not.

\paragraph{Interpretation.} None of the five probes---I/O shape, arrival burstiness, prefix caching (no-hit), variable-length arrivals, and prefix caching (real hits)---overturn the main finding. The headline underutilization penalty across the 7-point $\lambda$ sweep is a structural property of the continuous-batching scheduler, not an artifact of the I/O ratio, arrival process, length distribution, or APC state used to measure it.

\subsection{Measurement Stability}\label{sec:stability}

A natural reviewer question for any single-run benchmarking study is whether the reported spread is a structural property of the system or an artifact of measurement noise. We address this with a focused repeat-run campaign on the C2 configuration (Llama 3.1 8B FP8 on a single H100 NVL, 1xGPU deployment). We re-ran the random-token workload at four offered-load points $\lambda \in \{1, 10, 50, 100\}$ with three independent repeats each (12 runs total, distinct seeds per repeat), holding all other knobs fixed. The driver targets a $\geq60$\,s steady-state measurement window: prompt count is $60\cdot\lambda$ clamped to $[500, 6000]$ and warmup is $30\cdot\lambda$ floored at 100, producing $\{500,600,3000,6000\}$ prompts and $\{100,300,1500,3000\}$ warmups for $\lambda\in\{1,10,50,100\}$ respectively. At $\lambda{=}50$ and $\lambda{=}100$ the server is queue-limited and only $\sim$2{,}100 and $\sim$2{,}020 of the prompted requests complete within the measurement window; the CV computation uses the completed-request statistics reported by the harness. The vLLM server process stays resident across all 12 runs---a single server instance, so allocator and scheduler state are reused---which is consistent with how an operator would run back-to-back measurements and is the regime we are claiming stability for. Table~\ref{tab:stability-cv} reports the coefficient of variation (CV) across repeats on throughput, effective cost, and TTFT$_{p50}$.

\begin{table}[h]
\centering
\footnotesize
\setlength{\tabcolsep}{4pt}
\caption{C2 repeat-run measurement stability, based on $n{=}3$ independent repeats per $\lambda$ (12 runs total). Throughput and effective cost reproduce to within $0.31\%$ CV across the three repeats at every tested arrival rate. TTFT$_{p50}$ is similarly tight at under-saturated loads ($\lambda{=}1,10$) but widens at saturation-adjacent loads ($\lambda{=}50$), reflecting queue-tail sensitivity to arrival randomness rather than instrument error. $C_{\mathrm{eff}}$ CV values are near-identical to throughput CV because GPU-hour cost is a constant multiplier.}
\label{tab:stability-cv}
\begin{tabular}{rrrrrr}
\toprule
$\lambda$ & \multicolumn{2}{c}{Throughput (tok/s)} & \multicolumn{2}{c}{$C_{\mathrm{eff}}$ (\$/MTok)} & TTFT$_{p50}$ \\
(rps) & mean & CV\% & mean & CV\% & CV\% \\
\midrule
1   & 255.4  & 0.01 & 7.5920 & 0.01 & 0.33 \\
10  & 2501.8 & 0.31 & 0.7750 & 0.31 & 0.22 \\
50  & 7576.6 & 0.11 & 0.2559 & 0.11 & 8.73 \\
100 & 7419.5 & 0.12 & 0.2613 & 0.12 & 3.24 \\
\bottomrule
\end{tabular}
\end{table}

\paragraph{Takeaway.} The economic signal this paper reports---a $17.5\times$ to $36.3\times$ variation in $C_{\mathrm{eff}}$---is roughly four orders of magnitude larger than the measurement noise floor on $C_{\mathrm{eff}}$ when compared on the natural unit (cost-per-token in dollars): the between-run standard deviation on $C_{\mathrm{eff}}$ at $\lambda{=}1$ is \$$0.0004$/MTok against a cross-configuration spread of $\sim$\$15/MTok. Stated as a CV, reproducibility is $\leq0.31\%$ on $C_{\mathrm{eff}}$ at every tested $\lambda$. \emph{Assuming} the C2 noise profile transfers to other configurations---a simplification, since MoE routing dispatch and TP=2 collectives may carry different run-to-run variance---propagating the $0.31\%$ CV would imply approximate 95\% CIs of $[17.3, 17.7]\times$ for Mixtral FP16 and $[35.9, 36.7]\times$ for Qwen FP8. We report these as \emph{indicative} rather than measured; per-configuration repeat sweeps on the MoE headline configurations are the clean way to confirm them, and are noted as future work. The wider TTFT$_{p50}$ CV at $\lambda{=}50$ is not a stability concern for the cost claims but a real secondary observation: tail latency becomes queue-tail-limited once the scheduler is saturation-adjacent, and a single run reports one realization of that tail. Operators who care about TTFT SLOs near saturation should plan for run-to-run spread at the percent level on tail latency, even when throughput and cost are reproducible to within $\sim0.3\%$.

\subsection{Cross-Hardware Validation (A100 80GB PCIe)}\label{sec:cross_hardware}

A second obvious reviewer question is whether the load-driven spread we report is a property of H100 NVL specifically---its HBM3 bandwidth, its native FP8 tensor cores, its NVLink fabric---or a structural property of the offered-load $\lambda$ axis that would reproduce on a different GPU family. We repeat the core sweep on a second, architecturally distinct platform: NVIDIA A100 80GB PCIe (no native FP8, Gen4 PCIe in place of NVLink, 1.94\,TB/s HBM2e vs.\ H100's 3.9\,TB/s HBM3), at a list-price rate of \$3.67/hr/GPU on Azure's \texttt{NC\_A100\_v4} family (roughly $1.90\times$ cheaper than H100 NVL per GPU). We run the identical measurement harness---same prompt corpus, same arrival-rate ladder $\lambda \in \{1,5,10,25,50,100,200\}$, same two-gate health check, same steady-state window---across six paired configurations: Llama 3.1 8B FP16 and FP8 (vLLM, TP=1), Qwen3-30B-A3B FP16 and FP8 (vLLM, TP=1), Mixtral 8x7B FP16 (vLLM, TP=2), and Llama 3.1 8B FP16 (SGLang, TP=1). A third Mixtral condition---TP=4---was added on a 4$\times$A100 node to exercise a tensor-parallel dimension the 2$\times$H100 apparatus could not. This TP=4 run (M3) is reported separately in Result~4 and is excluded from the six-configuration 7.0--11.4$\times$ spread quoted above; its own spread of 12.3$\times$ is discussed in context. An eighth A100 configuration---Mixtral 8x7B FP16 on SGLang TP=2 (M2)---is present in the public 56-run corpus but excluded from Table~\ref{tab:cross_hardware_spread} because the SGLang engine did not scale on this hardware: throughput flatlined near 647~tok/s across every $\lambda$, consistent with a PCIe-interconnect bottleneck on TP=2 Mixtral that did not manifest on the NVLink-connected H100. The vLLM TP=2 row (M1) is reported in its place.

\begin{figure*}[!tp]
\centering
\includegraphics[width=0.95\textwidth]{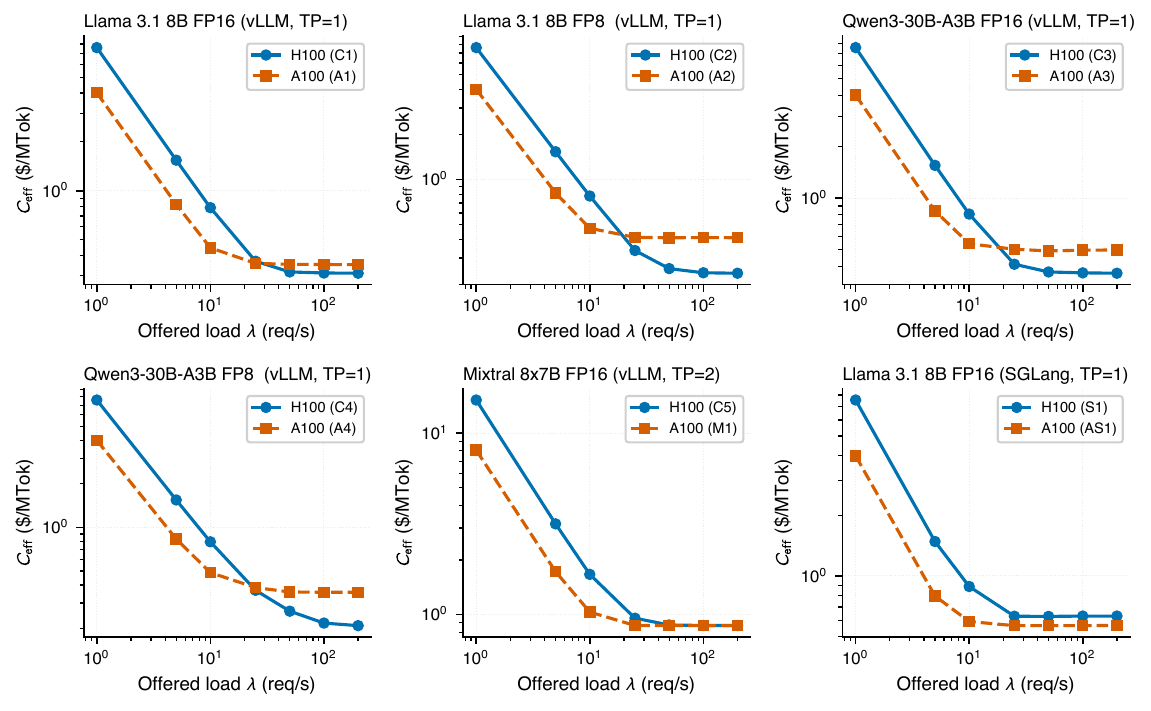}
\caption{Effective cost $C_{\mathrm{eff}}$ vs.\ offered load $\lambda$ on H100 NVL (solid blue) vs.\ A100 80GB PCIe (dashed orange), log-log, for six paired configurations. The saturation cliff reproduces on A100; the absolute spread is narrower (7.0--11.4$\times$ vs.\ 17.5--36.3$\times$) because A100's lower peak throughput and lower hourly rate compress the numerator-denominator ratio. At low $\lambda$ (near-idle), A100's lower hourly rate makes it cheaper per-million-tokens; above each panel's crossover (typically $\lambda\gtrsim 25$ for dense/FP8, later for Mixtral~TP=2), H100's throughput advantage wins despite the premium hourly rate.}
\label{fig:cross_hardware}
\end{figure*}

\paragraph{Result 1: the load-driven spread reproduces.} Figure~\ref{fig:cross_hardware} plots $C_{\mathrm{eff}}$ vs.\ $\lambda$ on log-log axes for each paired configuration, H100 (solid blue circles) overlaid on A100 (dashed orange squares). Every panel shows the same characteristic cliff: a steep drop from $\lambda{=}1$ to $\lambda{\approx}25$, followed by a plateau as the engine saturates. Quantitatively, the A100 data exhibits a 7.0--11.4$\times$ per-configuration spread (Table~\ref{tab:cross_hardware_spread}), narrower than the 17.5--36.3$\times$ observed on H100. Two mechanisms explain the compression: (i) the A100's lower peak throughput brings the saturation plateau closer to the idle-edge cost, and (ii) the A100's 47\% lower hourly rate proportionally shrinks the numerator of $C_{\mathrm{eff}}$ at idle. The \emph{shape} of the curve---a cliff at low $\lambda$ followed by saturation---is hardware-independent; the \emph{absolute magnitude} of the spread tightens on the cheaper, slower accelerator. This is the portable finding: any calculator that omits $\lambda$ and assumes a fixed utilization will be wrong by close to an order of magnitude on A100 and by over an order of magnitude on H100.

\begin{table}[h]
\centering
\scriptsize
\setlength{\tabcolsep}{3pt}
\caption{Per-configuration $C_{\mathrm{eff}}$ at saturation ($C_{\min}$, \$/MTok) and spread ($C_{\max}/C_{\min}$, unitless $\times$ multiplier) on both hardware families. Cost-per-token at saturation is within a factor of two between platforms for every configuration---and on Mixtral TP=2 they round to the same \$0.87/MTok at two decimal places (H100 \$0.8706 vs.\ A100 \$0.8688, a 0.2\% difference rather than a fundamental equality)---while the spread is always narrower on A100.}
\label{tab:cross_hardware_spread}
\begin{tabular}{@{}lrrrr@{}}
\toprule
 & \multicolumn{2}{c}{H100 NVL} & \multicolumn{2}{c}{A100 PCIe} \\
\cmidrule(lr){2-3}\cmidrule(lr){4-5}
Config. & $C_{\min}$ & spr. & $C_{\min}$ & spr. \\
\midrule
Llama 8B FP16 (vLLM)   & 0.31 & 24.4 & 0.35 & 11.4 \\
Llama 8B FP8 (vLLM)    & 0.24 & 31.9 & 0.41 & \phantom{0}9.8 \\
Qwen 30B FP16 (vLLM)   & 0.36 & 20.9 & 0.49 & \phantom{0}8.2 \\
Qwen 30B FP8 (vLLM)    & 0.21 & 36.3 & 0.36 & 11.3 \\
Mixtral FP16 TP=2 (vLLM)& 0.87 & 17.5 & 0.87 & \phantom{0}9.3 \\
Llama 8B FP16 (SGLang) & 0.63 & 12.0 & 0.57 & \phantom{0}7.0 \\
\bottomrule
\end{tabular}
\end{table}

\paragraph{Result 2: the FP8 MoE-first pattern reproduces on the architecture, but inverts on dense when the hardware lacks native FP8.} On H100 NVL, FP8 quantization cuts $C_{\min}$ by 23\% on Llama (0.31 $\to$ 0.24), 42\% on Qwen3-30B-A3B (0.36 $\to$ 0.21), and 40\% on Mixtral TP=2 (0.87 $\to$ 0.52). On A100 80GB PCIe---which lacks native FP8 tensor cores on its SM80 silicon and therefore runs FP8 kernels through a software-emulated path---FP8 \emph{worsens} Llama's saturation cost (0.35 $\to$ 0.41, $+17\%$) while still \emph{improving} Qwen's (0.49 $\to$ 0.36, $-27\%$). The architectural signal---that MoE models benefit from FP8 more than dense models do---survives the hardware change. What does not survive is the unconditional assumption that ``FP8 is cheaper'': on an accelerator without native FP8, the kernel emulation overhead dominates the memory-bandwidth savings for a dense 8B model, and the operator pays a small penalty for the quantization. Any planning framework that claims FP8 is a free lunch needs an explicit hardware caveat; this paper's framework already accommodates that caveat because $Q$ is a first-class input to $C_{\mathrm{eff}}(H, M, Q, \lambda, L)$.

\paragraph{Result 3: the active-parameters-beat-total-parameters claim survives the hardware change.} On H100 NVL the paper's headline ordering at saturation is Qwen3-30B-A3B FP8 (3B active, \$0.21/MTok) cheaper than Llama 3.1 8B FP8 (8B active, \$0.24/MTok)---active parameters, not total model size, drive the saturation cost. On A100 80GB PCIe the ordering holds: Qwen3-30B-A3B FP8 at \$0.36/MTok is cheaper than Llama 3.1 8B FP8 at \$0.41/MTok, and on the FP16 side Qwen3-30B-A3B at \$0.49/MTok is only modestly above Llama at \$0.35/MTok despite activating $3.75\times$ fewer parameters per token. Two hardware families now support the same claim: when a sparse-MoE model's active-parameter count is small relative to a dense competitor, the sparse-MoE model will reach a competitive cost-per-token at saturation, even carrying 3--6$\times$ more total weight in VRAM. Operators choosing between architectures should budget on active parameters.

\paragraph{Result 4: the TP=2 vs.\ TP=4 inversion on Mixtral.} An additional A100 configuration on a 4$\times$A100 node lets us probe a tensor-parallel scaling question the H100 apparatus could not. On A100, Mixtral 8x7B at TP=2 (engine: vLLM) saturates at \$0.8688/MTok; at TP=4 the same model saturates at \$1.3016/MTok---approximately $1.50\times$ more expensive per token (49.8\% at a single run per TP; a repeat-stability sweep on TP=4 is deferred to camera-ready) despite doubling the GPU count. Throughput does scale with TP (A100 TP=2 peaks at 2{,}347 tok/s, TP=4 at 3{,}133 tok/s, a 1.33$\times$ gain) but the 2$\times$ GPU-cost multiplier overwhelms the sub-linear throughput gain. This is a concrete instance of the paper's general claim that \$/MTok is decided by the denominator, not the numerator: adding GPUs to a configuration that already has sufficient capacity inflates $C_{\mathrm{eff}}$ even though it increases peak throughput. Cost-aware operators should pick the smallest TP that fits the model and the SLO, and scale horizontally via replicas rather than vertically via TP.

\paragraph{Takeaway.} The framework's central prediction---that $C_{\mathrm{eff}}$ moves by nearly an order of magnitude across $\lambda$ on identical hardware---reproduces on a second, architecturally distinct hardware family. The absolute spread tightens when moving to cheaper, slower silicon; the FP8-advantage pattern depends on whether the silicon has native FP8 kernels; the active-parameters-dominate claim survives. The cross-hardware exercise is, in effect, a falsification test: if H100 were a special case, the A100 curves would be flat. They are not. The recommendation to measure $C_{\mathrm{eff}}$ against \emph{one's own traffic and hardware} is now supported by data on two architecturally distinct hardware families.

% ============================================================
% 6. DISCUSSION
% ============================================================
\section{Discussion}\label{sec:discussion}

\subsection{Why Existing Calculators Fail}

Every calculator we surveyed---15+ of them---makes the same mistake: it treats GPU utilization as an input parameter the user is trusted to guess (or quietly assumes 100\%). Utilization is not an input. It is the \emph{output} of the interaction between arrival rate, model architecture, quantization, and SLO. A calculator that accepts utilization as a field is asking its users to solve the paper they came to the calculator to avoid solving. That is why the headline error is an order of magnitude and not, say, 2$\times$: the entire class of tools has the causal direction of its own input backwards.

\subsection{Practical Implications}

Beyond the calculators themselves, the offered-load axis is largely absent from how self-hosted inference is budgeted in practice: capacity plans and infrastructure reviews reason about GPU-hour and per-token rates as fixed inputs, leaving the variable that empirically moves realized cost the most---offered load---out of the cost model entirely. Restoring it as a first-class input, the framework enables three practical decisions:
\begin{enumerate}[leftmargin=*,topsep=2pt,itemsep=1pt]
    \item Given a target workload profile, what is the realistic cost of self-hosting?
    \item At what traffic volume does self-hosting become cheaper than a specific API tier?
    \item For a hybrid strategy, where should the routing boundary be set between self-hosted and API-served traffic?
\end{enumerate}

\paragraph{The penalty is also a provisioning signal, not only a calculator error.} The largest multipliers in this paper occur at offered loads (notably $\lambda{=}1$ rps) at which a cost-aware operator would not hold a dedicated accelerator idle: the structural remedy for a deployment sitting deep in the cost cliff is to \emph{change the provisioning}---consolidate tenants, scale replicas to the offered load, coalesce requests, or fall back to a serverless endpoint below the crossover---not merely to read a more accurate number. The framework supports both readings: an accurate $C_{\mathrm{eff}}(\lambda)$ is the \emph{diagnostic} that tells an operator they are over-provisioned, and the offered-load axis is the dimension along which autoscaling and consolidation act. This also explains why the minute-level dispersion in \S\ref{sec:live_validation} is a transient of a fixed deployment rather than a budget line: once the fleet size is itself elastic, cost-per-token at an instantaneous $\lambda$ is an attribution, not a monthly rate.

\noindent Contribution (3) connects directly to cost-aware routing systems. A routing controller that dynamically switches between self-hosted inference and API fallback needs an accurate real-time cost model---precisely what this framework provides.

\subsection{The Asymmetric Token Pricing Advantage}

A structural advantage of self-hosting that existing analyses consistently overlook: commercial API providers employ \textit{asymmetric token pricing}, charging substantially more for output tokens than input tokens. GPT-5.5 charges \$5.00/M input vs.\ \$30.00/M output~\cite{openai_pricing2025}; Claude Sonnet 4.6 charges \$3.00/M input vs.\ \$15.00/M output~\cite{anthropic_pricing2025}; Gemini 3.1 Pro charges \$2.00/M input vs.\ \$12.00/M output~\cite{google_pricing2026}. Output tokens are priced 5--6$\times$ higher because generation is compute-intensive and sequential---each token requires a full forward pass.

Self-hosted models incur identical GPU infrastructure cost regardless of whether tokens are input (parallel prefill) or output (sequential decode). This asymmetry has a concrete implication: the effective API cost for generation-heavy workloads is far higher than aggregate per-token pricing suggests. Self-hosting, by contrast, imposes no separate input-token and output-token tariff---though the input/output mix still moves the realized cost per output token through the prefill/decode balance (\S\ref{sec:sensitivity}), so a like-for-like comparison prices the API at its asymmetric per-token rates and the self-hosted deployment at its measured $C_{\mathrm{eff}}$ under the same workload shape.

As an illustrative back-of-the-envelope (not a measured benchmark), a code-generation pipeline with 100-token inputs and 500-token outputs ($\frac{5}{6}$ of tokens are output) pays approximately $(100 \times \$5 + 500 \times \$30) / 600 \approx \$25.80$/MTok aggregate at GPT-5.5 list pricing---well above the nominal \$5.00/M input rate typically quoted. The same tokens served on self-hosted Qwen3-30B-A3B FP8 carry no such output-token premium: its \$0.209/MTok saturation cost (measured at the 512:256 shape) bills input and output tokens at the same GPU-time rate, though the realized figure still shifts with workload shape (\S\ref{sec:sensitivity}). The asymmetric pricing effect amplifies the self-hosting advantage precisely for the workloads---code generation, long-form writing, agentic tool use---where enterprises are most likely to consider infrastructure investment.

Our framework currently models aggregate token throughput; a natural extension is to separate input and output token economics and incorporate the asymmetric pricing structure into crossover analysis.

\subsection{The Serverless-vs-Dedicated Fallacy}
\label{sec:serverless_fallacy}

A recurring question from infrastructure decision-makers is: \emph{``Why not just use a serverless API at \$X per million tokens?''} On its face, this appears to be a straightforward cost comparison. It is not. Published serverless per-token prices are \emph{unconditioned on any production service-level objective}. Commercial providers do not publish enforceable per-request guarantees for time-to-first-token or inter-token latency at the p99 tail; best-effort availability is the operative contract. A dedicated self-hosted deployment that must meet a production SLO (for example, TTFT p99 $\leq$ 300\,ms, TPOT p99 $\leq$ 50\,ms) operates in a different cost regime entirely, because provisioning headroom to absorb arrival-rate variance without breaching the tail is precisely what drives $C_{\mathrm{eff}}$ away from $C_{\mathrm{sat}}$.

Concretely: under our six-phase live workload (1--50\,rps), the best minute-level effective cost a Mixtral-8x7B FP16 deployment on 2$\times$H100 achieves is \$8.23/MTok (Table~\ref{tab:live_demo}, ``Best'' column, observed at the 50\,rps saturation minute)---still roughly 9.5$\times$ the theoretical saturation floor ($C_{\mathrm{sat}}{=}\$0.87$) because the averaging window covers the full arrival-rate traversal including idle sub-intervals. An operator pinned below saturation to preserve TTFT p99 headroom would sit strictly worse than this figure. The published commercial API price for a comparable class of model is often quoted at \$12--\$15 per million output tokens (the lower reference tiers of \S\ref{sec:crossover}). The two numbers are not comparable: one reflects a best-effort stateless endpoint with no latency SLA, the other reflects a dedicated deployment that has committed to a p99 latency contract. Choosing between them based on headline per-token price alone is a category error, analogous to comparing spot cloud instance pricing against reserved-instance pricing without acknowledging availability and preemption differences.

The framework presented in this paper makes this mismatch visible. An organization with a production SLO should compute $C_{\mathrm{eff}}$ at the arrival rate its SLO permits (\Cref{tab:slo_operating} works this through for a fixed example SLA), \emph{then} compare to serverless pricing while explicitly acknowledging that the serverless path surrenders tail-latency control. Our companion tool, \texttt{vllm-cost-meter}, gates its API-comparison feature behind an \texttt{-{}-accept-slo-mismatch} flag for exactly this reason: the comparison is meaningful only when the user has consciously accepted that serverless offers no SLO counterpart.

\subsection{Using This Framework Operationally}
\label{sec:operational_framework}

The framework is designed to be actionable by a single infrastructure owner in a three-step workflow.

\textbf{Step 1 --- Deploy your chosen configuration.} Select the model, quantization, and hardware that match your deployment constraints. The benchmark corpus released with this paper (the six H100 configurations C1--C6, plus the A100 cross-hardware and sensitivity sweeps; \S\ref{sec:artifact}) provides a starting reference, but is not a substitute for on-hardware measurement under your own workload shape.

\textbf{Step 2 --- Load-test under your own workload and SLO.} Run a sweep (for example, with \texttt{vllm bench serve}, \texttt{inference-perf}, \texttt{LLMPerf}, or \texttt{GenAI-Perf}\cite{inferenceperf2025,llmperf2024,genaiperf2024}) using your real prompt-length and output-length distribution, your expected arrival pattern, and your declared TTFT/TPOT/E2EL SLO. Record the highest arrival rate $\lambda$ at which your SLO is preserved; this is the \emph{goodput} of your deployment, distinct from the raw saturation $\Theta_{\max}$ reported here.

\textbf{Step 3 --- Plug observed goodput into the calculator.} Feed the goodput-at-SLO into the $C_{\mathrm{eff}}$ formula (Eq.~\ref{eq:ceff}) to obtain the effective cost of your deployment under your actual operating contract. Compare that number---not $C_{\mathrm{sat}}$, and not the raw $\Theta_{\max}$ from a no-SLO benchmark---to any alternative pricing.

This workflow inverts the usual procurement order. Rather than picking a price point and then discovering latency consequences, the operator declares the SLO first and derives the cost second. The \texttt{vllm-cost-meter} tool supports the third step directly: its live scrape of vLLM's Prometheus endpoint surfaces observed goodput in real time against user-declared SLO budgets, and computes $C_{\mathrm{eff}}$ continuously as arrival rate varies.

\subsection{The Open-Source Tool}

\texttt{vllm-cost-meter} is the concurrency-aware replacement for every calculator critiqued in Section~\ref{sec:related}.

\paragraph{Why a meter, not a calculator.} This is a deliberate design choice, not a missing feature. A calculator asks its user to supply the model's true maximum throughput on their hardware---precisely the SLO-aware, carefully measured quantity this paper argues is almost never obtained correctly, and that in practice is guessed or copied from a vendor benchmark. A meter inverts the dependency: by scraping the live server's own Prometheus metrics it \emph{derives} the operating-point throughput from observed traffic, so the hard-to-get denominator is measured rather than assumed. We therefore release an instrument that reads ground truth instead of a form that propagates a guess. The pre-computed curves for the six configurations characterized here serve as a bounded lookup for those exact deployments; we intentionally do \emph{not} extrapolate them into a general-purpose calculator, because a calculator populated with presets for unmeasured models would reproduce the same heuristic-preset, utilization-naive error (cf.\ the preset-based tool in \S\ref{sec:related}) that this paper exists to correct. A community calculator seeded by crowd-sourced \emph{meter} outputs---real observations, never presets---is a natural future direction.

Three things ship today:
\begin{itemize}[leftmargin=*,itemsep=2pt,topsep=2pt]
  \item \textbf{Benchmark runner.} One command sweeps vLLM across a chosen arrival-rate grid and emits a structured CSV.
  \item \textbf{Cost estimator.} Given a target $\lambda$, interpolates the empirical curve to produce a utilization-adjusted \$/MTok and a crossover point against any API price the user supplies.
  \item \textbf{Live dashboard.} Scrapes vLLM's \texttt{/metrics} Prometheus endpoint and plots $C_{\mathrm{eff}}$ in real time as arrival rate shifts.
\end{itemize}
Pre-computed curves from this paper's 42~runs ship in the repository; a practitioner running one of these benchmarked configurations can obtain a corrected cost estimate in under a minute without running a single benchmark, while unmeasured deployments require the on-hardware sweep of \S\ref{sec:operational_framework}.

\subsection{Live Validation}
\label{sec:live_validation}

To validate the framework beyond controlled benchmarks, we deployed \texttt{vllm-cost-meter} on the same H100~NVL hardware alongside live vLLM inference servers running all six benchmark configurations (C1--C6). The tool is an \emph{objective live instrument}: on every tick it scrapes vLLM's Prometheus \texttt{/metrics} endpoint, computes the live $C_{\mathrm{eff}}$ from observed generation tokens per second and declared hardware cost, and surfaces the instantaneous offered-load/cost relationship. A six-phase enterprise workload simulation ramped request rates from 1 to 50~rps and back, exercising the full range of arrival rates that a production deployment would traverse across a day.

Table~\ref{tab:live_demo} summarizes the results. In one hour of real traffic the same deployment traversed a 653$\times$ cost swing between its cheapest and most expensive \emph{minutes}---which we present as an \emph{illustration} of minute-window dispersion, not as a reproducible validation datum (per the caveat below). This is an extreme ratio: it contrasts the idle-valley instantaneous minute against the peak-load minute, not a time-weighted mean. Mixtral~FP16 peaked at \$5{,}379 per million tokens during the idle valley (when only a handful of tokens were billed against the full GPU-hour) and dropped to \$8.23 at 50~rps. Averaged over the hour the effective cost is dominated by the high-$\lambda$ minutes; the 653$\times$ figure is best read as a stress-test artifact of minute-window cost attribution near idle---a near-singular ratio that drops as the window lengthens---rather than as a monthly billing multiplier. We quote it to make the minute-level dispersion concrete; the underlying per-minute Prometheus ticks are not checked into the public corpus (Table~\ref{tab:live_demo} records the aggregated Best/Worst columns), so the 653$\times$ should be read as a directional signal rather than a reproducible data point. Any cost number quoted without a $\lambda$ attached---even averaged---is meaningless to a degree most practitioners badly underestimate.

The live validation demonstrates two things. First, the concurrency-aware framework produces actionable real-time estimates: the same deployment traces a recognizable $C_{\mathrm{eff}}$ curve as $\lambda$ varies, matching the shape predicted by the benchmark sweep. Second, the underutilization penalty is not merely a theoretical artifact of offline benchmarking---it manifests immediately in live operation whenever $\lambda$ falls below the arrival rate at which the engine saturates. The curve the tool surfaces is the same curve an operator would trace by re-running the benchmark sweep on their own workload; the tool makes this visible without a separate measurement pass. The validation we do claim is one of \emph{agreement}: the live, Prometheus-derived $C_{\mathrm{eff}}$ traces the same offered-load curve, at the same operating points, as the offline benchmark sweep on identical hardware---confirming that the meter reports the cost structure an operator would otherwise have to reconstruct by re-running the sweep.

\begin{table*}[!htbp]
\centering
\caption{Live validation: effective cost (\$/MTok) observed during six-phase workload simulation (1--50\,rps) on H100~NVL. 32{,}401 requests, 100\% success.}
\label{tab:live_demo}
\small
\setlength{\tabcolsep}{5pt}
\begin{tabular}{@{}llrrr@{}}
\toprule
\textbf{Config} & \textbf{Precision} & $C_{\mathrm{sat}}$ & \textbf{Best} & \textbf{Worst} \\
\midrule
C1: Llama-8B (1 GPU) & FP16 & 0.31 & 1.74 & 3{,}496 \\
C2: Llama-8B (1 GPU) & FP8  & 0.24 & 1.79 & 226 \\
C3: Qwen-30B (1 GPU) & FP16 & 0.36 & 1.63 & 202 \\
C4: Qwen-30B (1 GPU) & FP8  & 0.21 & 1.60 & 353 \\
C5: Mixtral (2 GPU)   & FP16 & 0.87 & 8.23 & 5{,}379 \\
C6: Mixtral (2 GPU)   & FP8  & 0.52 & 6.24 & 472 \\
\bottomrule
\end{tabular}
\end{table*}

\subsection{Scope and Future Work}

We deliberately scope the headline sweep to synthetic fixed-length workloads on a single H100~NVL node---with a targeted A100~80GB~PCIe validation sweep (\S\ref{sec:cross_hardware})---for reproducibility and to avoid exposing proprietary production data. The main benchmark sweep is run with prefix caching and speculative decoding disabled, establishing a conservative performance floor and cost upper bound; our sensitivity probes (\S\ref{sec:sensitivity}) partially relax these assumptions, enabling real prefix-cache hits and varying the arrival process and I/O shape. Production deployments with prefix caching, speculative decoding, and workload-specific tuning would achieve lower effective costs, making the absolute costs reported here upper bounds rather than typical-case estimates. Several extensions are natural: validating with variable-length conversational traces (e.g., ShareGPT, LMSYS-Chat-1M~\cite{lmsys_chat_1m}), incorporating SLO constraints as a first-class optimization target rather than an observed output, modeling diurnal traffic patterns and autoscaling economics, and extending to additional hardware generations (B200, MI300X) and inference engines (SGLang, TensorRT-LLM). The framework's structure is designed to accommodate these extensions, though each would require its own empirical validation to confirm the shape of $U(\lambda, L)$ under the new conditions.

\subsection{Limitations}\label{sec:limitations}

Our benchmarks use synthetic fixed-length workloads with uniform 512-token inputs and 256-token outputs; production workloads with variable-length conversational distributions will exhibit different utilization curves. We test three model architectures (dense, ultra-sparse MoE, sparse MoE); additional architectures (e.g., state-space models, encoder-decoder) may show different cost behaviors. The headline vLLM numbers are conducted on H100 NVL; we add a cross-hardware validation sweep on A100 80GB PCIe (\S\ref{sec:cross_hardware}) to rule out single-hardware confounding, and the load-driven spread reproduces there, though still-older hardware (V100) or newer generations (B200, MI300X) would shift the curves further and are left as future work. The headline numbers use a single engine (vLLM); engine choice preserves the qualitative cliff but shifts its \emph{magnitude}---the same Llama 3.1 8B FP16 configuration spreads 24.4$\times$ on vLLM versus 12.0$\times$ on SGLang (Table~\ref{tab:cross_hardware_spread}), as a lower $\Theta_{\max}$ compresses the idle-to-saturation ratio---so the absolute multiplier should be read as engine-conditional. We vary offered request rate while observing latency as an output rather than constraining it as an SLO input, leaving SLO-constrained optimization for future work. We do not model multi-tenant GPU sharing, which can improve utilization for providers serving multiple customers. GPU pricing changes frequently; the framework's value is in the methodology, not specific dollar amounts.

% ============================================================
% 7. CONCLUSION
% ============================================================
\section{Conclusion}\label{sec:conclusion}

The headline number is 17.5--36.3$\times$. On identical H100 hardware, holding model, precision, and GPU allocation fixed and varying only the offered request rate, effective cost per token moves by 17.5--36.3$\times$ within a single configuration; across all six configurations and the full load range it spans \$0.21 to \$15.25 per million output tokens. That spread is not a tail case---it is the dominant term in the cost equation, and it is invisible to every production cost calculator we surveyed.

Three practical implications. \textbf{First:} if an infrastructure plan is based on a per-token calculator, the number can be wrong by more than an order of magnitude---always in the direction that makes self-hosting look cheaper than it is. \textbf{Second:} on the three architectures tested (n=3; dense, ultra-sparse MoE, sparse MoE), active parameter count and memory-access patterns, not total parameter count, appear to drive saturation economics: Qwen3-30B-A3B~FP8 is cheaper per token at saturation than Llama~3.1~8B~FP8, inverting the size-based intuition. The same ordering survives the A100 replication in \S\ref{sec:cross_hardware}, so two silicon families now support the claim, but it still rests on three architectures; broader validation---in particular a larger dense model (e.g., Llama~3.1~70B) and a third MoE sparsity ratio---is required before the pattern can be treated as an architectural-scaling law. \textbf{Third:} FP8 is not a dense-model optimization---on our three models the two MoE architectures benefit $2.2$--$2.4\times$ more than the dense one (+69 to +74\% vs.\ +31\% peak throughput), a structural signal that should shift quantization priorities pending wider confirmation.

The tool is \texttt{vllm-cost-meter}. It scrapes a running vLLM server's Prometheus endpoint and surfaces the real cost curve in real time. If today's answer to ``what does a million tokens cost us?'' is a single number, replace it with a curve before the next infrastructure review.

\subsection{Data and Artifact Availability}\label{sec:artifact}
The companion \texttt{vllm-cost-meter} tool and the raw per-run CSVs are released under an MIT license at \url{https://github.com/pChitral/vllm-cost-meter}: the 140-run core corpus---the 42-run H100 vLLM headline sweep, the 42-run H100 SGLang companion sweep, and the 56-run A100 80GB PCIe cross-hardware sweep---together with the I/O-shape, prefix-caching, and burstiness sensitivity sweep (\S\ref{sec:sensitivity}) and the measurement-stability campaign (\S\ref{sec:stability}). All saturation points, $C_{\mathrm{eff}}$ spreads, underutilization penalties, and stability CVs in \S\ref{sec:results} are re-derivable directly from these CSVs; the variable-length and real-prefix-hit probes of \S\ref{sec:sensitivity} are summarized in-text. A version-tagged Zenodo DOI will be minted at camera-ready. The repository is intended as a community substrate for extending the $\lambda$-axis sweep to additional hardware, model architectures, and SLO-bounded regimes.

% ============================================================
% REFERENCES
% ============================================================
\clearpage % flush any deferred double-column floats BEFORE the bibliography
\FloatBarrier
\balance
\bibliographystyle{plain}

\begin{thebibliography}{20}

\bibitem{pan2025breakeven}
G.~Pan, V.~Chodnekar, A.~Roy, and H.~Wang.
\newblock A cost-benefit analysis of on-premise large language model deployment: Breaking even with commercial LLM services.
\newblock {\em arXiv preprint arXiv:2509.18101}, 2025.

\bibitem{slam2024}
C.~Irugalbandara, A.~Mahendra, R.~Liyanage, et al.
\newblock Scaling down to scale up: A cost-benefit analysis of replacing {OpenAI}'s {LLM} with open source {SLM}s in production.
\newblock In {\em IEEE ISPASS}, 2024.

\bibitem{vidur2024}
A.~Agrawal, N.~Kedia, J.~Mohan, et al.
\newblock {VIDUR}: A large-scale simulation framework for {LLM} inference.
\newblock In {\em MLSys}, 2024.

\bibitem{melange2024}
T.~Griggs, X.~Liu, J.~Yu, et al.
\newblock {M\'{e}lange}: Cost efficient large language model serving by exploiting {GPU} heterogeneity.
\newblock {\em arXiv preprint arXiv:2404.14527}, 2024.

\bibitem{sageserve2025}
S.~Jaiswal et al.
\newblock {SageServe}: Optimizing {LLM} serving on cloud data centers with forecast aware auto-scaling.
\newblock {\em arXiv preprint arXiv:2502.14617}, 2025.

\bibitem{erdil2025}
E.~Erdil.
\newblock Inference economics of language models.
\newblock {\em arXiv preprint arXiv:2506.04645}, 2025.

\bibitem{wingpt2025}
B.~Zhuang, J.~Qiao, M.~Liu, et al.
\newblock Beyond benchmarks: The economics of {AI} inference.
\newblock {\em arXiv preprint arXiv:2510.26136}, 2025.

\bibitem{infermax2024}
K.~Kim, J.~Li, K.~Hong, and A.~Ailamaki.
\newblock Faster {LLM} inference using {DBMS}-inspired preemption and cache replacement policies ({INFERMAX}).
\newblock {\em arXiv preprint arXiv:2411.07447}, 2024.

\bibitem{vllm2023}
W.~Kwon, Z.~Li, S.~Zhuang, et al.
\newblock Efficient memory management for large language model serving with {PagedAttention}.
\newblock In {\em SOSP}, 2023.

\bibitem{nvidia_tco2025}
{NVIDIA}.
\newblock {LLM} inference benchmarking: How much does your {LLM} inference cost?
\newblock {\em NVIDIA Developer Blog}, June 2025.
\newblock \url{https://developer.nvidia.com/blog/llm-inference-benchmarking-how-much-does-your-llm-inference-cost/}. Accessed 2026-04-23.

\bibitem{introl2026}
{Introl}.
\newblock Inference unit economics: The true cost per million tokens.
\newblock {\em Introl Blog}, February 2026.
\newblock \url{https://introl.com/blog/inference-unit-economics-true-cost-per-million-tokens-guide}. Accessed 2026-04-23.

\bibitem{acnicessc2025}
{acnicessc}.
\newblock {LLM} Inference {TCO} Calculator v2.4.
\newblock \url{https://acnicessc.github.io/llmcalc/}, 2025. Accessed 2026-04-23.

\bibitem{bentoml2025}
{BentoML}.
\newblock llm-optimizer: Benchmark and optimize {LLM} inference across frameworks.
\newblock \url{https://github.com/bentoml/llm-optimizer}, 2025. Accessed 2026-04-23.

\bibitem{semianalysis2025}
{SemiAnalysis}.
\newblock {InferenceX} (formerly {InferenceMAX}): Open-source continuous inference benchmarking.
\newblock \url{https://github.com/SemiAnalysisAI/InferenceX}, 2025. Accessed 2026-04-23.

\bibitem{guidellm2025}
{vLLM Project}.
\newblock {GuideLLM}: Evaluate {LLM} deployments for real-world inference.
\newblock \url{https://github.com/vllm-project/guidellm}, 2025. Accessed 2026-04-23.

\bibitem{qwen3_2025}
{Qwen Team}.
\newblock Qwen3 Technical Report.
\newblock {\em arXiv preprint arXiv:2505.09388}, 2025.

\bibitem{llama31_2024}
A.~Grattafiori, A.~Dubey, A.~Jauhri, et al.
\newblock The {Llama} 3 Herd of Models.
\newblock {\em arXiv preprint arXiv:2407.21783}, 2024.

\bibitem{mixtral_2024}
A.~Q.~Jiang, A.~Sablayrolles, A.~Roux, et al.
\newblock Mixtral of Experts.
\newblock {\em arXiv preprint arXiv:2401.04088}, 2024.

\bibitem{openai_pricing2025}
{OpenAI}.
\newblock {API} Pricing.
\newblock \url{https://developers.openai.com/api/docs/pricing}, 2026. Accessed 2026-06-09.

\bibitem{anthropic_pricing2025}
{Anthropic}.
\newblock Claude {API} Pricing.
\newblock \url{https://claude.com/pricing}, 2026. Accessed 2026-06-09.

\bibitem{google_pricing2026}
{Google}.
\newblock Gemini {API} Pricing.
\newblock \url{https://ai.google.dev/gemini-api/docs/pricing}, 2026. Accessed 2026-06-09.

\bibitem{inferenceperf2025}
{Kubernetes SIG Serving}.
\newblock inference-perf: GenAI inference performance benchmarking tool.
\newblock \url{https://github.com/kubernetes-sigs/inference-perf}, 2025. Accessed 2026-04-23.

\bibitem{llmperf2024}
{Anyscale}.
\newblock {LLMPerf}: A tool for benchmarking LLMs.
\newblock \url{https://github.com/ray-project/llmperf}, 2024. Accessed 2026-04-23.

\bibitem{genaiperf2024}
{NVIDIA}.
\newblock {GenAI-Perf}: Benchmark generative AI models.
\newblock \url{https://github.com/triton-inference-server/perf_analyzer}, 2024. Accessed 2026-04-23.

\bibitem{splitwise2024}
P.~Patel, E.~Choukse, C.~Zhang, et al.
\newblock Splitwise: Efficient generative {LLM} inference using phase splitting.
\newblock In {\em ISCA}, 2024.

\bibitem{distserve2024}
Y.~Zhong, S.~Liu, J.~Chen, et al.
\newblock {DistServe}: Disaggregating prefill and decoding for goodput-optimized large language model serving.
\newblock In {\em OSDI}, 2024.

\bibitem{sarathiserve2024}
A.~Agrawal, N.~Kedia, A.~Panwar, et al.
\newblock {Taming Throughput-Latency Tradeoff in LLM Inference with Sarathi-Serve}.
\newblock In {\em OSDI}, 2024.

\bibitem{sglang2024}
L.~Zheng, L.~Yin, Z.~Xie, et al.
\newblock {SGLang}: Efficient execution of structured language model programs.
\newblock In {\em NeurIPS}, 2024.

\bibitem{lmsys_chat_1m}
L.~Zheng, W.-L. Chiang, Y.~Sheng, et al.
\newblock {LMSYS-Chat-1M}: A large-scale real-world {LLM} conversation dataset.
\newblock In {\em ICLR}, 2024.

\bibitem{orca2022}
G.-I.~Yu, J.~S.~Jeong, G.-W.~Kim, et al.
\newblock {Orca}: A distributed serving system for transformer-based generative models.
\newblock In {\em OSDI}, 2022.

\bibitem{mooncake2024}
R.~Qin, Z.~Li, W.~He, et al.
\newblock {Mooncake}: A {KVCache}-centric disaggregated architecture for {LLM} serving.
\newblock {\em arXiv preprint arXiv:2407.00079}, 2024.

\bibitem{lmcache2025}
Y.~Liu, Y.~Cheng, J.~Yao, et al.
\newblock {LMCache}: An efficient {KV} cache layer for enterprise-scale {LLM} inference.
\newblock {\em arXiv preprint arXiv:2510.09665}, 2025.

\bibitem{llumnix2024}
B.~Sun, Z.~Huang, H.~Zhao, et al.
\newblock {Llumnix}: Dynamic scheduling for large language model serving.
\newblock In {\em OSDI}, 2024.

\bibitem{burstgpt2024}
Y.~Wang, Y.~Chen, Z.~Li, et al.
\newblock {BurstGPT}: A real-world workload dataset to optimize {LLM} serving systems.
\newblock {\em arXiv preprint arXiv:2401.17644}, 2024.

\bibitem{servegen2025}
Y.~Xiang, X.~Li, K.~Qian, et al.
\newblock {ServeGen}: Workload characterization and generation of large language model serving in production.
\newblock {\em arXiv preprint arXiv:2505.09999}, 2025.

\end{thebibliography}

\end{document}